\documentclass[acmtog,nonacm]{acmart}
\acmSubmissionID{119}
\usepackage{multirow}
\usepackage{booktabs} 
\usepackage{natbib}
\usepackage{graphicx}
\usepackage{float}
\usepackage{enumitem}
\usepackage{bbding}
\usepackage{soul}
\usepackage{bm}
\usepackage{xspace}
\usepackage{array}
\usepackage{overpic}
\usepackage{color}
\usepackage[normalem]{ulem}
\usepackage[ruled]{algorithm2e} 

\SetAlFnt{\small}
\SetAlCapFnt{\small}
\SetAlCapNameFnt{\small}
\SetAlCapHSkip{0pt}

\setcopyright{rightsretained}
\acmJournal{TOG}
\acmYear{2023} \acmVolume{42} \acmNumber{4} \acmMonth{8} \acmArticle{1} \acmPrice{15.00}
\acmDOI{10.1145/3592093}

\newcommand{\modelname}{HACK\xspace}

\begin{document}
\title{\modelname: Learning a Parametric Head and Neck Model for High-fidelity Animation}


\author{Longwen Zhang}\authornote{Equal contributions.}
\orcid{0000-0001-8508-3359}
\affiliation{%
 \institution{ShanghaiTech University and Deemos Technology Co., Ltd.}
 \city{Shanghai}
 \country{China}}
\email{zhanglw2@shanghaitech.edu.cn}

\author{Zijun Zhao}\authornotemark[1]
\affiliation{%
 \institution{ShanghaiTech University and Deemos Technology Co., Ltd.}
 \city{Shanghai}
 \country{China}}
\email{zhaozj2022@shanghaitech.edu.cn}

\author{Xinzhou Cong}\authornotemark[1]
\affiliation{%
 \institution{ShanghaiTech University and Deemos Technology Co., Ltd.}
 \city{Shanghai}
 \country{China}}
\email{congxzh2022@shanghaitech.edu.cn}

\author{Qixuan Zhang}
\affiliation{%
 \institution{ShanghaiTech University and Deemos Technology Co., Ltd.}
 \city{Shanghai}
 \country{China}}
\email{zhangqx1@shanghaitech.edu.cn}

\author{Shuqi Gu}
\affiliation{%
 \institution{ShanghaiTech University}
 \city{Shanghai}
 \country{China}}
\email{gushq@shanghaitech.edu.cn}

\author{Yuchong Gao}
\affiliation{%
 \institution{ShanghaiTech University}
 \city{Shanghai}
 \country{China}}
\email{gaoych@shanghaitech.edu.cn}

\author{Rui Zheng}
\affiliation{%
 \institution{ShanghaiTech University}
 \city{Shanghai}
 \country{China}}
\email{zhengrui@shanghaitech.edu.cn}

\author{Wei Yang}
\affiliation{%
 \institution{Huazhong University of Science and Technology}
 \city{Wuhan}
 \country{China}}
\email{weiyangcs@hust.edu.cn}

\author{Lan Xu}\authornote{Corresponding author.}
\affiliation{%
 \institution{ShanghaiTech University}
 \city{Shanghai}
 \country{China}}
\email{xulan1@shanghaitech.edu.cn}

\author{Jingyi Yu}\authornotemark[2]
\affiliation{%
 \institution{ShanghaiTech University}
 \city{Shanghai}
 \country{China}}
\email{yujingyi@shanghaitech.edu.cn}

\renewcommand\shortauthors{Zhang, L. et al}

\begin{abstract}
Significant advancements have been made in developing parametric models for digital humans, with various approaches concentrating on parts such as the human body, hand, or face. Nevertheless, connectors such as the neck have been overlooked in these models, with rich anatomical priors often unutilized. In this paper, we introduce HACK (Head-And-neCK), a novel parametric model for constructing the head and cervical region of digital humans. Our model seeks to disentangle the full spectrum of neck and larynx motions, facial expressions, and appearance variations, providing personalized and anatomically consistent controls, particularly for the neck regions.
To build our HACK model, we acquire a comprehensive multi-modal dataset of the head and neck under various facial expressions. We employ a 3D ultrasound imaging scheme to extract the inner biomechanical structures, namely the precise 3D rotation information of the seven vertebrae of the cervical spine. We then adopt a multi-view photometric approach to capture the geometry and physically-based textures of diverse subjects, who exhibit a diverse range of static expressions as well as sequential head-and-neck movements.
Using the multi-modal dataset, we train the parametric HACK model by separating the 3D head and neck depiction into various shape, pose, expression, and larynx blendshapes from the neutral expression and the rest skeletal pose. We adopt an anatomically-consistent skeletal design for the cervical region, and the expression is linked to facial action units for artist-friendly controls. We also propose to optimize the mapping from the identical shape space to the PCA spaces of personalized blendshapes to augment the pose and expression blendshapes, providing personalized properties within the framework of the generic model.
Furthermore, we use larynx blendshapes to accurately control the larynx deformation and force the larynx slicing motions along the vertical direction in the UV-space for precise modeling of the larynx beneath the neck skin. HACK addresses the head and neck as a unified entity, offering more accurate and expressive controls, with a new level of realism, particularly for the neck regions. This approach has significant benefits for numerous applications, including geometric fitting and animation, and enables inter-correlation analysis between head and neck for fine-grained motion synthesis and transfer.

\end{abstract}

%
%

\ccsdesc[500]{Computing methodologies~Mesh models}

\keywords{Head and neck modeling, Anatomical model, Facial expressions, Neck animation, Parametric learning}

\begin{teaserfigure}
    \setlength{\abovecaptionskip}{3pt}
    \centering
    \includegraphics[width=1\textwidth]{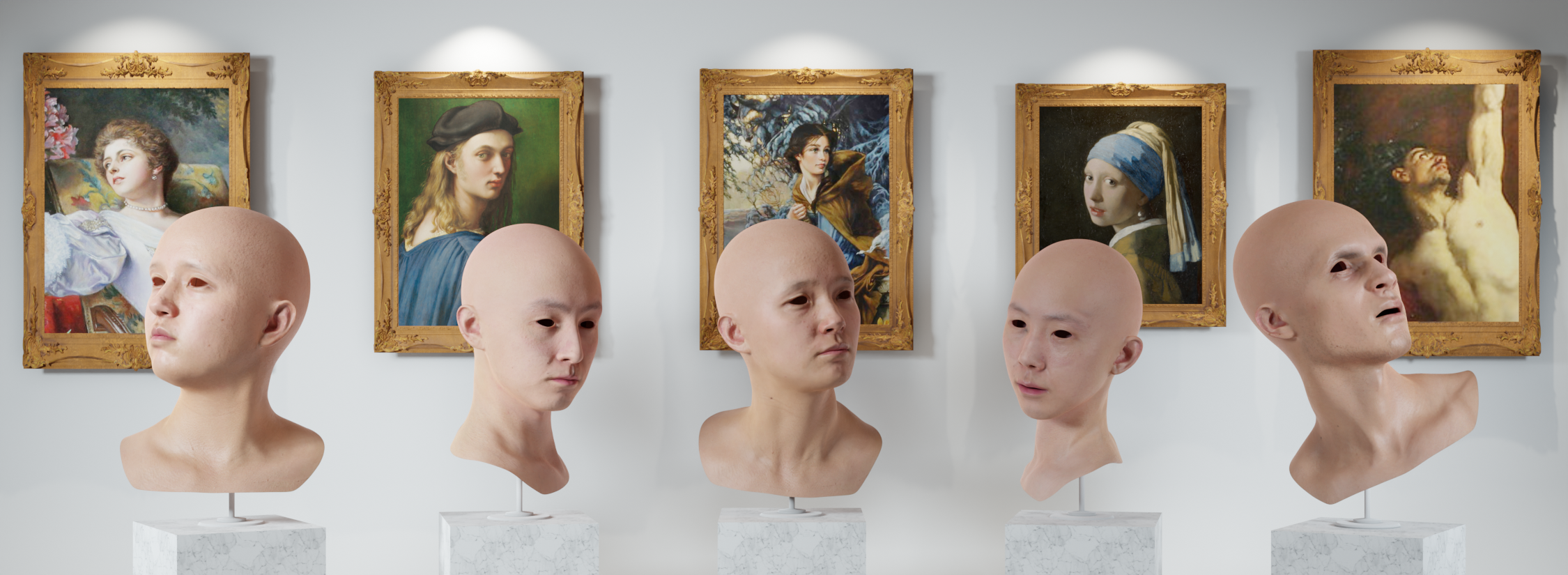}
    \caption{
{
we present HACK (Head-And-neCK), a novel parametric model for constructing the cervical region of digital humans. By combining rich physically-based appearance modeling and inner anatomical structures, HACK achieves more accurate and expressive results than existing head and neck models.
}
    }
    \label{fig:teaser}
\end{teaserfigure}

\maketitle

\section{Introduction}

Realism in human modeling goes beyond surface-level characteristics. Anatomical structures, physical attributes, and their induced motions capture unique and nuanced personalities, encompassing their inner thoughts, emotions, and experiences. In constructing digital humans, tremendous efforts have been focused on modeling the skin, skeleton, and muscle components of faces \cite{Sculptor,2001facemusclemodel}, hands \cite{2017mano,Li2021PIANO}, bodies \cite{loper2015smpl} that exhibit more obvious features. In contrast, body parts that serve as connectors have been largely overlooked. A significant example is the human neck which connects the jaw, head, and shoulder, allowing the head and face to move in a more natural way. The neck is often visible in a variety of poses and angles, and it greatly impacts the overall appearance of a digital human character. For example, the prominent jugular vein on the neck of Michelangelo's monumental David sculpture vividly illustrates the adrenaline-fueled situation of David, before facing the giant Goliath.

Convincing neck movements further add subtlety and realism to facial performances as well as enable nonverbal communication and social interaction. For instance, we humans turn to an unconscious habit of air swallowing when overly anxious. In fact, the idiosyncrasies of the head and neck are the defining characteristics of various human characters, real or virtual, physical or digital. However, so far very few attempts have been made to model human necks~\cite{Luo2013anatomyneck,NeckTVCG2021}, let alone a generic parametric model for the face and neck as a whole. The challenges are multifold. Anatomically the neck is made up of bones, muscles, and other tissues that are difficult to model. Therefore, to allow for a wide range of movement including flexion, extension, rotation, and lateral bending, it is critical to accurately model the joints in accord with the head. Further, the neck is a flexible and deformable structure under facial motions. Yet by far very limited datasets are available that simultaneously capture the facial and neck geometry under a rich variety of motions.

The majority of existing parametric models build upon 3DMM~\cite{blanz1999morphable,Brunton20143DMM,Booth20173DMM} to model the human head ~\cite{Sculptor,yang2020facescape,wang2022faceverse}, as large as the whole front face and as small as individual facial components such as eyeballs, teeth, lips, etc ~\cite{EyeModel2016,TeethModel2016,LipModel2016}.
A few attempts ~\cite{LearningFaceCVPR2020,NeckTVCG2021} aim to extend the parametric model to the neck region. They unanimously employ a highly simplified model, e.g., using a single empirical joint, to represent the kinematics between head and neck, without considering more sophisticated anatomical priors of inner structures such as the cervical spine. Physiologically, the animation of human head and neck is an intricate orchestration, from rich facial expressions coordinated with neck movements to nuanced deformation of the neck shape caused by sliding larynx beneath the skin. Without accurately incorporating these anatomical priors, brute-force extensions of the parametric model can often lead to unconvincing and sometimes biomechanically absurd results as shown in Fig.~\ref{fig:comparison}.

Early 3DMM-based models~\cite{LearningFaceCVPR2020,yang2020facescape} also tend to retain the statistical properties of the underlying 3D scan dataset based on an inherent low-rank approximation and therefore are insufficient to represent local and high-frequency surface details. More recent data-driven approaches ~\cite{yenamandra2021i3dmm,MoFaNeRFeccv2022,hong2021headnerf,MoRFtog2022,giebenhain2022nphm} adopt neural rendering techniques to provide more personalized and realistic results across various identities. A drawback though is that the latest neural models cannot readily support the existing CG production pipeline to conduct either rendering or controllable editing. The most recent trend~\cite{2020Dynamic,LearningFaceCVPR2020,AuthenticTOG2022,NeRFBlendShape} is to explicitly employ more personalized characteristics into the generic parametric models. However, they rely on tedious subject-specific training or network inference to obtain personalized facial assets, sacrificing the compact and efficient controls of 3DMM-based models.

In this paper, we present HACK (Head-And-neCK), a novel parametric model for constructing the cervical region of digital humans. By combining the outer physically-based appearance and inner anatomical structures (i.e., the cervical spine that is composed of seven vertebrae), HACK tackles the full spectrum of neck and larynx motions, offering more personalized and anatomically-consistent controls with a new level of realism (see Fig.~\ref{fig:teaser}). As a parametric model analogous to previous blendshape-based techniques~\cite{loper2015smpl,FLAME:SiggraphAsia2017}, HACK is differentiable, computationally efficient, and compatible with existing CG engines.

The first step in building HACK is data collection for effective model training. We first acquire a comprehensive multi-modal dataset that covers both internal anatomical structures and external appearances of the head and neck under facial expressions.
To extract the cervical spline as biomechanical priors, we use the portable 3D ultrasound imaging (US) system~\cite{Chen2020chaosheng} to obtain the sonography scans of the neck regions of individuals. Such a solution is radiation-free and cost-effective and the process has obtained IRB approval.
We then ask experienced radiologists to label 3D landmarks on the seven vertebrae of the cervical spine and subsequently extract their anatomically-consistent rotation information with regard to the skull and the external neck geometry.
To correlate the inner structures of the neck with its outer appearance, we conduct physically-based scanning using a photometric scanning solution ~\cite{2000lightstage,Debevec2012LSX,Zhang2022VideoDriven}. Specifically, we employ the multi-view photometric capture system to capture both the geometry and physically-based textures of head and neck regions.
We further conduct topology-consistent reconstructions to capture the physically-based head-and-neck attributes on subjects who perform a diverse set of static expressions as well as sequential motions.
Large-scale geometry variations can be effectively recovered in terms of 3D surfaces via 3D/4D scans.
Besides, we directly obtain the high-resolution 2D RGB images and normal maps to model small-scale and nuanced variations (e.g., movements of the larynx).

To learn the parametric HACK model, we follow a similar discipline as in the human face and body modeling~\cite{FLAME:SiggraphAsia2017,loper2015smpl} by separating the depiction of 3D head and neck into various shape, pose, expression, and larynx blendshapes from the neutral expression and the rest skeletal pose.
Specifically, to learn pose-dependent blendshapes, we adopt an anatomy-consistent skeleton of the neck that consists of 8 joints corresponding to the 7 cervical vertebrae and the head skull, respectively.
We also tie the expression blendshapes to the action units of the Facial Action Coding System (FACS)~\cite{2015FACS,li2010example,2020Dynamic} to provide a compact and artist-friendly expression control.
Instead of using the same generic blendshapes across identities similar to previous parametric models~\cite{FLAME:SiggraphAsia2017,loper2015smpl}, we further tailor schemes to learn personalized pose and expression blendshapes, by optimizing the generic mapping from the identical shape space to personalized blendshapes.
We show such a strategy significantly enhances HACK's capability of modeling personalized properties while maintaining the generalization across identities as a generic model.
Inspired by the recent work~\cite{NeckTVCG2021}, we further adopt the larynx blendshapes to control the larynx deformation on top of the larynx-removed neck under the rest pose, and subsequently force the larynx slicing motions along the vertical direction in the UV-space.
Such a strategy provides the disentanglement to anatomically mimic two kinds of muscles related to the larynx: one moves the vocal folds and hence changes the larynx's size while the other causes vertical slicing of the larynx beneath the neck skin.
Once trained, our HACK model is differentiable and compatible with standard CG software, and hence readily benefits a variety of applications like geometric fitting, animation, and visual inference.
Most importantly, our HACK model provides more accurate and expressive controls for spine-driven neck poses and larynx motions.
It also enables fine-grained analysis of the inter-correlation of head and neck, including:
(1) Faithful motion synthesis by leaning the temporal mapping from the head pose to the skeletal pose of the cervical spine, from facial expression to the larynx slicing.
(2) Biology-consistently transferring the head/neck motions from a human to another mammal i.e., the giraffe, since through evolution most mammals share the same skeletal structure of cervical spines.
To summarize, our main contributions include:
\begin{itemize}
    \setlength\itemsep{0em}

    \item {We present HACK, a generic parametric model that jointly considers human identity, facial expression, anatomy-inspired neck, and larynx motions, as well as physically-based appearance.}

    \item {We propose to jointly utilize the comprehensive inner biomechanical priors and external appearances during parameter learning, which provides personalized and anatomically-consistent controls for the neck regions.}

    \item {We make available our trained HACK model and showcase various applications to demonstrate its effectiveness, ranging from model fitting and inference, to motion synthesis and transfer.}

\end{itemize}

\begin{figure*}[t]
    \centering
    \includegraphics[width=\linewidth]{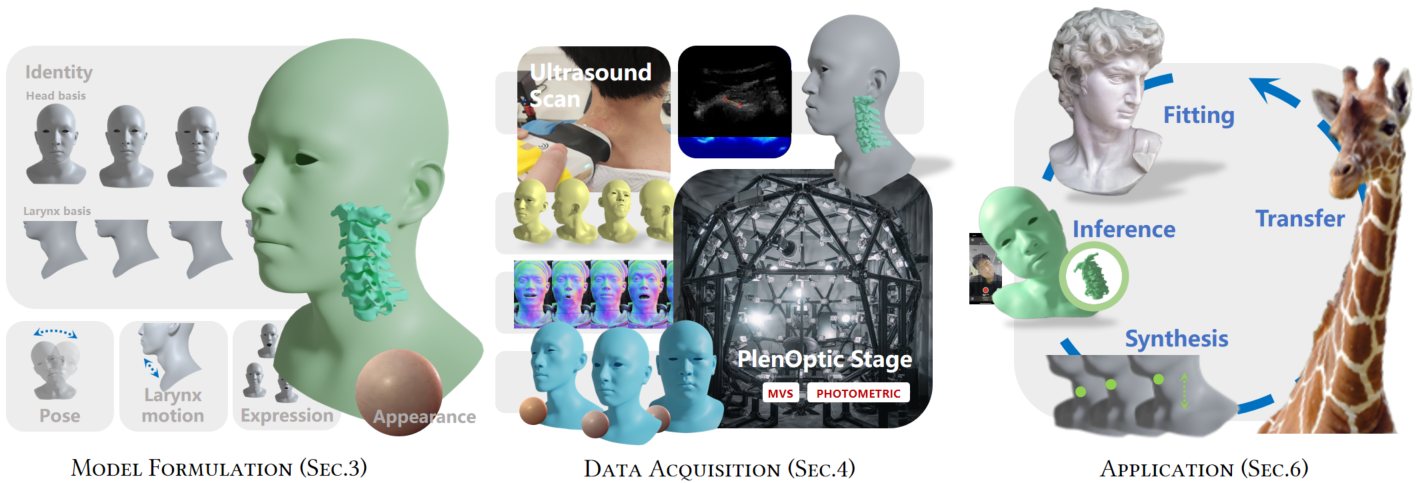}
    
    \vspace{-0.2cm}
    
    \caption{
Overview of \modelname. We first introduce the model formulation of HACK in Sec.~\ref{sec:ModelFormulation}, and the comprehensive multi-modal dataset that covers both internal anatomical structures and external appearances under facial expressions in Sec.~\ref{sec:data}. After building HACK, various applications are shown in Sec.~\ref{sec:application} to demonstrate its effectiveness.
    }
    \label{fig:overview}
\end{figure*}

\section{Related Works}

\paragraph{Parametric Head and Neck Models}

Parametric human modeling is characterized by a fixed number of parameters and a specific functional form, which makes modeling of the human body~\cite{2005SCAPE,loper2015smpl,Joo2018Adam}, hands~\cite{2017mano}, heads~\cite{2009parahead,FLAME:SiggraphAsia2017,Ploumpis2019parahead}, and faces~\cite{2009parahead,Brunton20143DMM,HuberP2016surrey3DMM,Booth20173DMM,Abrevaya20183DMM,2020york3DMM,smith2020mvsface,wang2022faceverse} relatively simple to work with and interpret.
Modeling and analysis of human faces are particularly important in the field as it plays a key role in many applications such as facial animation, virtual try-on, and face editing. 
Since \citet{blanz1999morphable} propose the first key ideas of general face representation, i.e., linear combinations of faces produce morphologically realistic faces, and separating facial shape and color to disentangle illumination, camera parameters and etc, many advances have been emerging~\cite{2009parahead, kim2018deep, FLAME:SiggraphAsia2017}.
We refer the readers to \citet{20193DMMsurvey} for a comprehensive survey on 3DMM. 
Though achieving high quality on the face region, most existing approaches pay little attention to the neck, which results in unrealistic neck models and animations. Some exceptions~\cite{FLAME:SiggraphAsia2017,EMOCA:CVPR:2021,loper2015smpl,STAR:2020,SUPR,Choi2022CGface} incorporate the neck modeling, which is still relatively simple and exhibit unrealistic deformations. Especially, \citet{NeckTVCG2021} and \citet{Li2013larynx} take the larynx into account and map 2D UV texture to the vertex displacement on 3D meshes, which produce good results.

A recent trend in human body modeling is to enforce anatomical constraints, several exemplary works include the modeling of the full human body \cite{Xu2020GHUM}, head \cite{Duan2013anatomyhead,Duan2015anatomyhead,Sculptor}, and hand models \cite{Li2021PIANO}. 
However, anatomical constraints of the neck region are rarely adopted in generic head and neck modeling \cite{Lee2006anatomyneck,Lee2009anatomyneck,Luo2013anatomyneck}. The reason is twofold: 1. anatomical constraints are good for the physical simulation of a specific subject but are difficult to generalize; 2. they lack details of external neck skin deformation. In this paper, we adequately consider anatomical constraints and construct a head and neck model with finer neck deformation.

Generating a realistic and convincing person-specific head-and-neck model requires tedious steps of manual intervention, including the construction of physically-based appearance attributes, the capture of fine facial movements and their migration \cite{LearningFaceCVPR2020,Thies2016head,Laine2017head}. These tasks generally require a lot of manual adjustments by professional artists, making the whole process too expensive to scale to a larger audience. Recently many learning-based approaches enable generic models that are highly robust for different lighting and motion \cite{FLAME:SiggraphAsia2017,DECA:Siggraph2021,Bao2021parametric,Raj2021generichead} or performer-specific high-quality geometry and appearance \cite{Gafni2020personalhead,Lattas2021personalhead,Gecer2019personalhead,Lombardi2019personalhead,Lombardi21} by processing datasets of different volumes. The former is able to learn person-specific blendshapes and pore-level dynamic physical materials \cite{2020Dynamic}, but generic blendshapes tend to lead to the loss of fine-grained expressions and dynamic physically-based textures. The latter can achieve very fine geometry and appearance for specific performers, and can even be re-lighted for animation and cross-performer driving. However, they mostly lack explicit control support and are difficult to deploy into traditional production processes.

\paragraph{Facial Animation}

The goal of 3D facial animation is to make the character's facial expressions and movements as realistic and believable as possible as if the character is a real human or creature. 
Typically, a 3D mesh is attached to a skeleton (or bones) and moves the bones to deform the mesh~\cite{wang2017realistic}. Instead, muscle-based animation uses physics-based models to simulate the movement and deformation of muscles in a 3D mesh. One example is the Facial Action Coding System (FACS), which annotates and codes facial expressions based on the movement of specific facial muscles \cite{ekman2002facial,mattar2015facs, mattar2017facs, jackson2017automatic}. Another example is the Facewarehouse model \cite{yang2020facescape,Deng2019facewarehouseRelated}, which represents the shape and appearance of a face using a set of blendshapes and a physics-based muscle simulation model.
Additionally, a face model can be driven by audio signals or video footage of a person's face. The Audio-Driven Facial Animation (ADFA) framework \cite{liu2019audio} uses deep learning to map between audio features and facial expression parameters, and the Audio-Driven Facial Animation System (ADFAS) \cite{fan2019audio} combines rule-based and data-driven techniques to synthesize realistic facial animations from audio signals.
2D Video-Driven Facial Animation (VDFA) systems combine optical flow and deep learning to synthesize facial animations from video input~\cite{guo2018video}, while 3D VDFA systems use morphable model fitting and facial expression synthesis to generate realistic facial animations~\cite{wang2020vdfa}.

\paragraph{Geometry and Appearance}

Capturing fine facial movements is an essential setup of head and neck modeling, and several advanced acquisition systems have already been proposed. Early approaches adopt single acquisition devices, such as laser scanners \cite{blanz2003face,phillips2008frgc,Levoy2000lazer} and structured light scanners \cite{Geng2011structuredlight}. Structured light systems capture at a single viewpoint and may not be sufficient to reconstruct a 3D geometry of the human face accurately. Multi-view stereo (MVS) can overcome this limitation by reconstructing a 3D geometry from 2D images of multiple views \cite{seitz2006mvs, wrobel2001mvs, goesele2006mvs}, where several successful works produce good results in recovering the human body \cite{Dou2016mvsbody, Zhang2022VideoDriven}, face \cite{klaudiny2012mvsface, lombardi2018mvs, Lombardi21, Chen2019mvsface,smith2020mvsface}, hand \cite{2017mano}, hair \cite{Nam2019mvshair}, etc. 

Photometric stereo is another prevalent technique for facial data acquisition \cite{jain2017dynamic, ma2006facial}. Instead of locating 3D points from pixel correspondences across multi-view images, photometric stereo systems capture images from single or sparse viewpoints under various lighting conditions and use the intensity variations to estimate the surface normals~\cite{Wang2020PSface, Zafeiriou2012PSface,villarini2017PSface}. Photometric stereo can be used to reconstruct the surface detail of a face, including wrinkles, texture, and color. 
The most well-known facial capture system using photometric stereo is the LightStage \cite{2000lightstage}, which has been widely used and modified in a large variety of fields in recent years~\cite{2005LS5,Debevec2012LSX,2006similarLS1,2011similarLS2}. The FaStage proposed in \cite{Zhang2022VideoDriven} extended the LightStage by combining multi-view reconstruction and photometric reconstruction to recover the dynamic geometry and physically-based texture of the performer in a motion sequence.

\paragraph{Anatomical Data Acquisition}

Exploiting anatomically prior information for human modeling has been a widely researched area, with great success being achieved through the use of Magnetic Resonance Imaging (MRI) and Computer Tomography (CT) scans. These techniques are widely used to reconstruct 3D models for evaluating bone and joint morphology and preoperative planning \cite{stephen2021MRICT, Touati2021MRICT}. CT scans provide a clear bone-soft tissue contrast but expose the subject to high doses of ionizing radiation. On the other hand, MRI is radiation-free, but it is more often used to scan soft tissue \cite{lacono2015MRI,Misaki2015MRIsoft} rather than bone structures \cite{Samim2021MRI}.

Ultrasound technology then is an alternative technique for capturing human bones in various applications \cite{Nelson1998US}. \citet{wang2018ultrasonic} used ultrasound range finding to capture detailed 3D models of the neck. \citet{Eerd2014USneck} scanned cervical spine specimens using an ultrasound probe that moves on a fixed track. In other studies, ultrasound waves have been used to track and analyze neck movements in various tasks such as head turning and tilting \cite{zhang2018ultrasonic, zhang2019ultrasonic}. Recently, \citet{Chen2020chaosheng} proposed a portable mobile 3D ultrasound scanning system, which we adopt to capture and reconstruct the interior structure of the cervical spine for anatomically prior information.

\section{Model Formulation}
\label{sec:ModelFormulation}

Here we introduce a novel parametric model, HACK, for jointly constructing the head and neck region of digital humans. As shown in Fig.~\ref{fig:overview}, our HACK model combines the rich observations from both physically-based appearance and inner anatomical structures. It achieves full-spectrum modeling of neck and larynx motions, providing personalized and anatomically-consistent controls. Analogous to the human head and body model~\cite{FLAME:SiggraphAsia2017,loper2015smpl}, our HACK models the depiction of 3D head and neck into various shape, pose, expression and larynx blendshapes from the neutral expression and the rest skeletal pose. Specifically, the general formulation of HACK is defined as follows:
\begin{equation}
\mathbf{HACK}(\bm{\beta},\bm{\psi},\bm{\theta},\eta,\tau, \bm{\alpha})=  \{  \mathbf{G}(\bm{\beta},\bm{\psi},\bm{\theta},\eta,\tau) ,\mathbf{A}(\bm{\alpha}) \},
\end{equation}
where $\mathbf{G}$ denotes the head-and-neck geometry, and $\mathbf{A}$ produces the physically-based appearance. 
$\bm{\beta}$, $\bm{\theta}$, $\bm{\psi}$ and $\bm{\alpha}$ are parameters to control the shape, pose, expression and appearance, respectively. 
Besides, we use $\eta$ and $\tau$ to control the larynx size and larynx slicing beneath the neck skin, so as to separate larynx motions from facial expressions for more subtle and realistic modeling (e.g., for handling swallowing). 
Specifically, the geometry model is obtained through a skinning process as follows:
\begin{equation}
   \mathbf{G}(\bm{\beta},\bm{\psi},\bm{\theta},\eta,\tau) = \mathbf{LBS} \big ( T(\bm{\beta},\bm{\psi},\bm{\theta},\eta,\tau), J(\bm{\beta}), \bm{\theta}, \mathcal{W} \big ),
   \label{eq:LBS}
\end{equation}
where $\mathbf{LBS}(\cdot)$ denotes the Linear Blend Skinning (LBS) function; $T(\cdot)$ is the person-specific mesh with corrective deformations ascribed to identity, pose, expression, and larynx parameters; $J(\bm{\beta})$ represents the joint location; $\mathcal{W}$ is the learned skinning weight of $\mathbf{LBS}(\cdot)$.
In stark contrast, we adopt an anatomy-consistent skeleton that consists of 8 joints corresponding to the 7 cervical vertebrae and the head skull. Hence, the joint position regressor $J(\cdot)$ infers the 8 person-specific biomechanical joint positions given identity parameter $\mathbf{\beta}$ for accurate skinning (Sec.~\ref{sec:HNSkeletonModel}). 

Then, the personalized template under rest pose is a linear combination of the universal template and various blendshapes:
\begin{align}
    T(\bm{\beta},\bm{\psi},\bm{\theta},\eta,\tau) = & \mathbf{\bar{T}} + \text{B}_\text{S}(\bm{\beta};\mathcal{S})+ \text{B}_\text{E}(\bm{\psi};\mathcal{E}_{\bm{\beta}})+ \text{B}_\text{P}(\bm{\theta};\mathcal{P}_{\bm{\beta}}) \nonumber \\
    & +L({\bm{\beta}}, \eta, \tau;\mathcal{L}).
    \label{eq:T-beforetheta}
\end{align}
Note that the universal template $\mathbf{\bar{T}}\in \mathbb{R}^{3N}$ is under the rest pose with mean shape, no expression, and larynx-removed. $\text{B}_\text{S}$ is the multiplication of orthonormal PCA of shape blendshapes $\mathcal{S}$ and $\bm{\beta}$. $\text{B}_\text{E}$ is the multiplication of expression blenshapes $\mathcal{E}_{\bm{\beta}}$ and $\bm{\psi}$.
$\text{B}_\text{P}$ is the multiplication of pose blendshapes $\mathcal{P}_{\bm{\beta}}$ and pose rotation matrix. Please refer to previous work~\cite{FLAME:SiggraphAsia2017,loper2015smpl} for detailed formulation. Yet, such unified blendshapes turn to retain the statistics of the whole dataset, thus losing personalized details. To this end, we propose to model the PCA spaces of the various person-specific $\mathcal{E}_{\bm{\beta}}$ and $\mathcal{P}_{\bm{\beta}}$, respectively, and subsequently learn an efficient mapping from the identity parameter $\bm{\beta}$ to their corresponding PCA spaces, so as to obtain person-specific facial motion traits (Sec.~\ref{sec:SEPbs}).

We further adopt the larynx deformation {$L(\bm{\beta}, \eta, \tau;\mathcal{L})$} following previous work~\cite{NeckTVCG2021}, which will be explained in Sec.~\ref{sec:larynxModeling}. Specifically, we first predict the larynx geometry on top of the larynx-removed neck by controlling its size according to $\eta$. Then, we formulate the larynx slicing motions along the vertical direction in the UV-space in terms of $\tau$.

\begin{figure}[t]
    \centering
    \includegraphics[width=\linewidth]{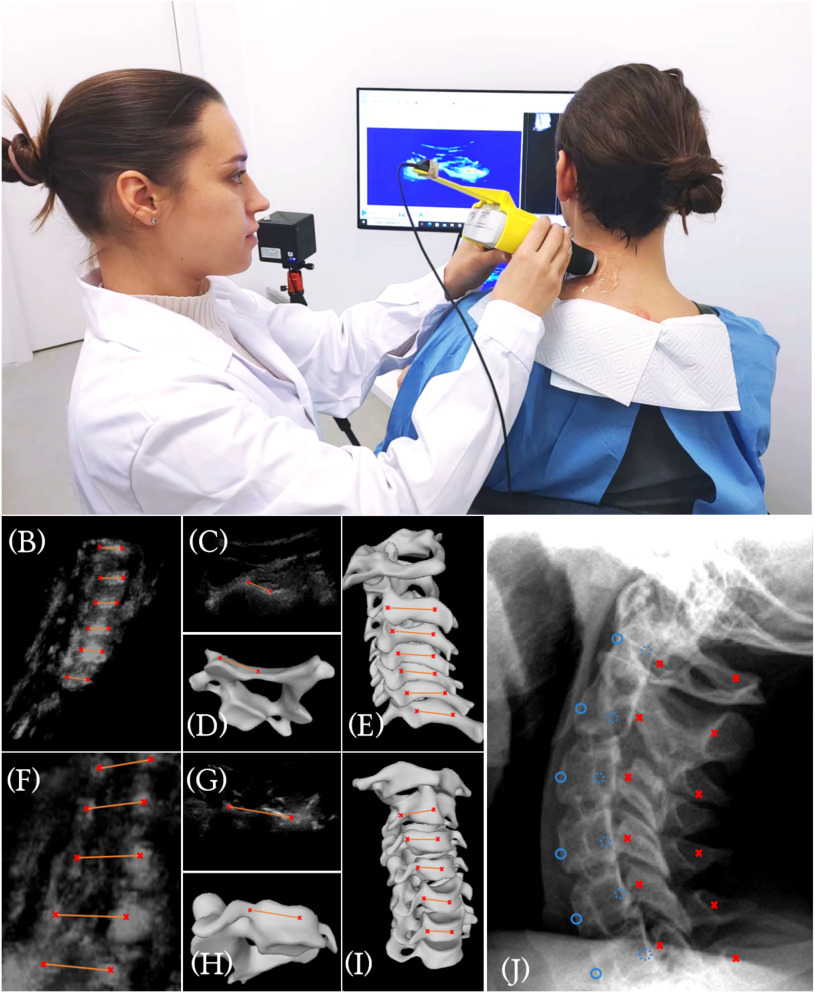}
    \caption{Ultrasound data acquisition pipeline and results.
    (A) A performer was scanned using the portable real-time 3D ultrasound imaging system. 
    (B) Reconstructed image from rear-left side scan and feature marker. 
    (C) 2D slice of rear-left side scan with feature marker.
    (D) Single vertebra mesh corresponding to rear-left side US scan.
    (E) Full cervical spine mesh corresponding to rear-left side US scan.
    (F) Reconstructed image from front-right side scan and feature marker. 
    (G) 2D slice of front-right side scan with feature marker.
    (H) Single vertebra mesh corresponding to front-right side US scan.
    (I) Full cervical spine mesh observed from the front-right side.
    (J) Feature marks on the sagittal human neck X-ray image, blue markers and red markers correspond to the front-right scan and rear-left scan, respectively.
        }
    \label{fig:ultrasonic}
\end{figure}

\subsection{Anatomy-aware Skeleton in HACK}
\label{sec:HNSkeletonModel}

An over-simplified skeleton, e.g., using a single empirical joint, to represent the kinematics between head and neck is insufficient and often leads to unconvincing animations. To this end, in our HACK model, we design an anatomy-inspired skeletal structure with 8 joints, which correspond to 7 cervical vertebrae (C1-C7) and the head skull (the apex of C1), respectively. These joints, denoted as c7-t1, c6-c7, c5-c6, c4-c5, c3-c4, c2-c3, c1-c2, and o-c1, are approximately located behind the intervertebral discs, and provide accurate kinematic control for nuanced neck motions.

To accurately determine the set of person-specific joint positions $\mathbf{J} \in \mathbb{R}^{3K}$, we resort to a portable 3D ultrasound imaging (US) system to obtain sonography scans with 3D information of the seven vertebrae, labeled by experienced radiologists (see Sec.\ref{subsec:USData}).
Accompanied by our anatomy-consistent skeleton design, we then optimize the regressor $J(\cdot)$ to predict joint positions from the rich ultrasound data in the rest pose.
Compared with the existing one-joint solutions, our HACK provides more accurate control over head and neck poses by accurately locating rotation centers and fully modeling the rotation of each cervical vertebra.

\subsection{More Personalized Expression/Pose Blendshapes}
\label{sec:SEPbs}
Recall that we follow the similar discipline of blendshapes as in human face
and body modeling~\cite{FLAME:SiggraphAsia2017,loper2015smpl}. Specifically, we adopt the orthonormal PCA of shape blendshapes, to correlate the shape parameter $\bm{\beta}$ with rich identity information. Instead of using the same generic expression and pose blendshapes that remain for all individuals, we further tailor schemes to employ more personalized properties into these blendshapes.
We propose to learn the extra mappings $\mathcal{M}_{E}$ and $\mathcal{M}_{P}$ from ${\bm \beta}$ to expression and pose blendshapes basis $\mathcal{E}_\beta$, $\mathcal{P}_\beta$ for novel identities, i.e., $\mathcal{M}_{E} ({\bm \beta}) \mapsto \mathcal{E}_\beta $ and $\mathcal{M}_{P} ({\bm \beta}) \mapsto \mathcal{P}_\beta$.

For expression deformations, we tie the expression blendshapes to the action units of FACS~\cite{2015FACS} for more artist-friendly expression controls, following previous work~\cite{2020Dynamic}. Specifically, for each captured subject with shape parameter $\bm{\beta}$, we obtain the corresponding expression blendshapes from the static captured scans with FACS expressions: $\mathcal{E}_{\bm{\beta}}=[\mathbf{E}_1^{\bm{\beta}},\dots,\mathbf{E}_{|\bm{\psi}|}^{\bm{\beta}}]$. Thus, all the individuals share the same latent structure for the expression parameters $\bm{\psi}$. To efficiently learn the mapping $\mathcal{M}_{E}$, we first formulate the PCA space of expression blendshapes from the set $\{{\mathcal{E}_{\bm{\beta}}}\}$ across various identities. We subsequently train the mapping network $\mathcal{M}_{E}$ as a shallow MLP to predict the PCA weights, hence generating the personalized expression blendshapes. 

For pose-dependent deformations, similar to previous work~\cite{loper2015smpl}, we employ the pose blendshapes $\mathcal{P}_{\bm{\beta}}$ with the rotation matrix interpreted from the skeletal parameter $\bm{\theta}$. Here we adopt an anatomy-aware skeletal design, where the $\bm{\theta}$ denotes the concatenated rotation vector of the joints for 7 cervical vertebrae and the head skull. 
Again, we obtain the person-specific $\mathcal{P}_{\bm{\beta}}$ for the performer with identity $\bm{\beta}$ from the captured dynamic sequential scans, and subsequently optimize the PCA space of pose blendshapes from  $\{{\mathcal{P}_{\bm{\beta}}}\}$. Analogous to $\mathcal{M}_{E}$, we adopt the same mapping network with shallow MLP to predict the PCA weights from the identity parameter. Such a strategy significantly improves the ability of our parametric model for generating more personalized controls.

\begin{figure*}
    \centering
    \includegraphics[width=\linewidth]{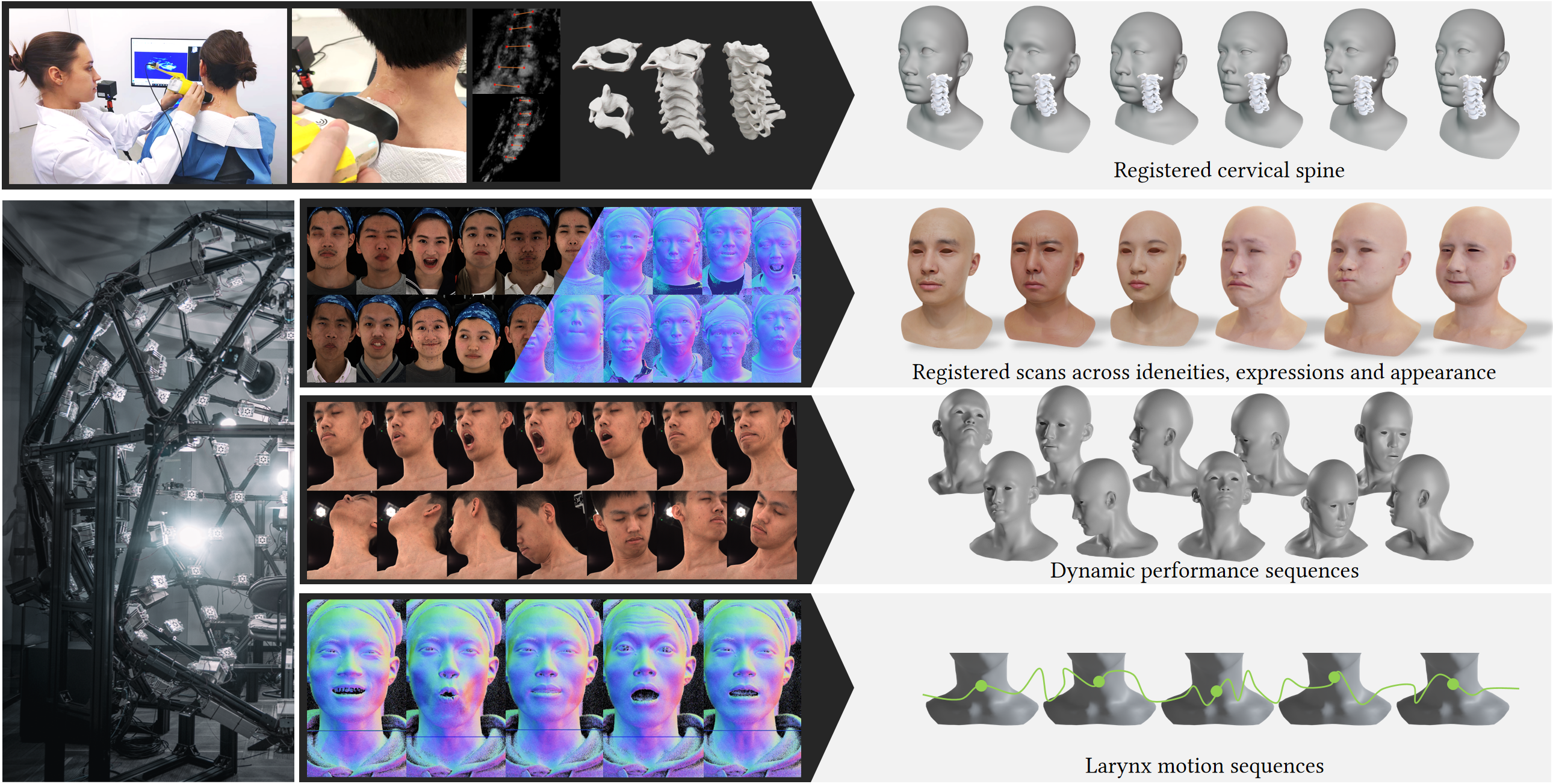}
    \caption{
Data processing pipeline.
Utilizing both the ultrasound imaging system and multi-view photometric capture system, we further process these multi-modal data for model learning. The processed data includes registered cervical spine joints, registered neutral meshes, solved personalized expressions, dynamic performance sequences, larynx motion sequences, and physically-based textures.
}
    \label{fig:dataset}
\end{figure*}

\subsection{Larynx Modeling}
\label{sec:larynxModeling}
Accurate geometric modeling of the larynx part is vital for realistic head and neck presentation, especially for actions with subtle larynx movement (e.g., swallowing and talking). Anatomically, the movement of the larynx is controlled by two groups of muscles, one of which moves the vocal folds and changes the size of the larynx, and the other moves the position of the larynx in the neck vertically. Hence, we model the larynx as vertex displacements added to the larynx-removed rest-pose mesh with shape change capability and constrained movement along the vertical direction of the neck,
which is also similar to~\cite{NeckTVCG2021}. 
$L(\bm{\beta}, \eta, \tau;\mathcal{L}): \mathbb{R}^{2 + |\bm{\beta}|} \mapsto \mathbb{R}^{3N}$ is the larynx shape function that maps identity $\bm{\beta}$, larynx size $\eta$ and position $\tau$ to vertex displacements to simulate the sliding effect beneath neck skin. 
\begin{equation}
    L(\bm{\beta}, \eta,\tau; \mathcal{L}) = \eta \cdot \sum_{i=1}^{|\bm{\beta}|} \beta_i \mathrm{L}_i(\tau),
\end{equation}
where $\mathcal{L} = [\mathrm{L}_1(\tau),\dots,\mathrm{L}_{|\bm{\beta}|}(\tau)]\in \mathbb{R}^{3N\times |\bm{\beta}|}$ is the larynx blendshape basis that is able to move vertically according to $\tau$. In actual implementation, we model the larynx geometric displacement in the UV space for convenience. Specifically, $\mathrm{L}_i (\tau) \in \mathbb{R}^{3 \times H \times W}$ is a larynx blendshape basis represented as a 2D image through texture atlasing and storing the vertex displacement into corresponding uv coordinate. Then the larynx function can be reformulated as: $L(\bm{\beta}, \eta,\tau; \mathcal{L}) = \{ \mathbf{v}(u, v) \}, \, \mathbf{v}(u, v) = \eta \cdot \sum_{i=1}^{|\bm{\beta}|} \mathrm{L}_i(u, v+\tau)$, where $\mathbf{v}(u, v)$ denotes the displacement of vertex $\mathbf{v}$ ascribe to larynx, which maps to the value at $u, v$ in $\mathrm{L}$. 
This formulation guarantees that the larynx moves vertically along the neck and hence increases robustness.

\section{Data Acquisition and Processing}
\label{sec:data}

In order to build HACK, a comprehensive and multi-modal dataset that covers both internal anatomical structures and the external appearance of the head and neck is essential for effective training. 
To acquire this dataset, we first introduce an ultrasound pipeline to construct biomechanical priors from sonography scans of the cervical spines (Sec.~\ref{subsec:USData}). 
Then, we utilize a multi-view photometric capture system to capture both the geometry and appearance of the head and neck across diverse facial expressions and sequential motions, where geometry variations are effectively recovered (Sec.~\ref{sec:mvs}).

\subsection{Neck Ultrasound Scanning}\label{subsec:USData}

To accurately extract the cervical spine as biomechanical priors, we employ a portable real-time 3D ultrasound imaging system~\cite{Chen2020chaosheng} to acquire sonography scans of the interior structure of the neck region, whose process includes scanning, annotating, and alignment. It is worth noting that our scanning pipeline is radiation-free and cost-effective, and the process has been approved by the IRB, which is demonstrated in Fig.~\ref{fig:ultrasonic}.

\paragraph{Ultrasound scan process}
During the ultrasound data collection process, the subject sits with their back against the chair to minimize body movement during scanning.
An ultrasound probe, equipped with an electromagnetic (EM) tracking sensor, is pressed against the subject's neck and moved from C1 to C7 along the cervical curve, as illustrated in Fig.~\ref{fig:ultrasonic}(A), or capturing continuous ultrasound images of the cervical spine. Before starting, an ultrasound coupling agent is applied to enable effective transmission of the ultrasound waves.
We scan the neck regions of each subject in the following order: rear-left, rear-right, front-left, and front-right. Fig.~\ref{fig:ultrasonic}(C) and Fig.~\ref{fig:ultrasonic}(G) present ultrasound images of a neck in the neutral pose, scanned in the rear-left and front-right directions, respectively.
 
\paragraph{Landmarks annotation}

To accurately locate the joints, we first reconstruct the 3D ultrasound volume of the neck and invite medical expertise to mark a set of pre-defined vertebra feature points. Specifically, the expertise annotates different points in ultrasound images captured from different neck regions. For instance, the points of interest of the rear-left side scans are the peak of the lamina and the bottom of the vertebra arch, while those of the front-right side scans are the end and center points of the vertebra body. 
Fig.~\ref{fig:ultrasonic}(B, C), (F, G) shows the annotated feature points for the rear-left, and front-right side scans respectively. And Fig.~\ref{fig:ultrasonic}(D, H, E, I) exhibit the corresponding feature points on the template cervical spine mesh.

\paragraph{Neck template alignment}

The final stage is to deform a standard cervical spine template to match each identity using annotated feature points and then put the deformed cervical spine into the model in the rest-pose. 
We first resize the template cervical spine mesh to fit with the real-world size ultrasound scan, i.e., we scale the vertebra in mesh to match the length in scans.
The vertebra is rigid, and its motion and deformation could be approximated as the translation and rotation of the whole vertebra. We need to estimate the rotation and position of the vertebra for accurate alignment~\cite{gycius}
, and we define the position of a vertebra as the center of the feature point pair on it. Then, the rotation is calculated through the line that connects the feature point pair. 
Applying the translation and rotation to the scaled template mesh, we can align the template with ultrasound data accurately, as exhibited in Fig.~\ref{fig:ultrasonic}(E, I, J) that the aligned template is consistent with the ultrasound scan. Further, we need to put the cervical spine template into the neck and head mesh in the rest pose. For this purpose, we ask the experienced radiologists to additionally annotate several markers on the outer skin surface on the ultrasound images, and subsequentially align the rest pose mesh with the neck template.

\paragraph{Anatomical-prior head and neck skeleton}

With the aligned cervical spine template in the rest pose mesh, we can finally define the joints of our head and neck skeleton model. We design the \modelname to have $K=8$ joints corresponding to the bottom points of 7 vertebrae (C1-C7) and the apex of C1. This setup yields 8 bone transformations denoted as c7-t1, c6-c7, c5-c6, c4-c5, c3-c4, c2-c3, c1-c2, and o-c1, that will serve as the foundation for skinning as in Eqn.~\ref{eq:LBS}.
By using sonography scans of the neck regions, we build biomechanical priors for HACK, which play an important role in further parametric learning and realistic animation.

\subsection{Multi-view Photometric Scanning} \label{sec:mvs}
To further correlate the inner structures of the neck with its outer appearance, and to achieve a diverse range of blendshapes for identity, expression, and pose, we conduct physically-based scanning using a photometric scanning solution. Specifically, we use a multi-view photometric capture system to acquire both the geometry and physically-based textures of the head and neck regions, ranging from a diverse set of static expressions to dynamic performance sequences. This allows us to effectively recover large-scale geometry variations and obtain detailed deformation of the head and neck, which is essential for building HACK and achieving realistic and nuanced movements.

\paragraph{Static head and neck capture}
\label{sec:static_scans}

To learn the shape and expression space, we utilize the multi-view photometric capture system to capture the static head and neck of various identities in rest pose with a neutral expression, including 274 females and 350 males, ranging from age 15 to 65, with different races, skin tones, and face shapes.

Besides neutral faces in rest pose, we further capture scans of subjects with various expressions (28 predefined expressions following the FACS standard~\cite{2015FACS}) for creating personalized expression blendshapes for each identity.
We conduct multi-view stereo reconstruction and register a template head and neck mesh with the reconstructed geometry following the registration pipeline as described in \citet{2020Dynamic}, under the supervision of densely painted markers on faces. Then, we invite several professional artists to remove the larynx from the registered template mesh for explicitly estimating the larynx shape and the subsequential blendshapes space construction.Additionally, we use the captured photometric data to recover the high-resolution physically-based textures pertaining to reflectance, including the diffuse, specularity, and normal maps, which are further used to build up the appearance space.  

\paragraph{Dynamic performance sequence capture}

To analyze the geometric deformation of head and neck-related poses, we capture the performing data under continuously changing poses of various identities.
In each captured clip, we require the performer to perform the following actions that will affect the head and neck muscles sequentially:
\begin{itemize}
    \item rest pose, neutral expression;
    \item rest pose, swallow once;
    \item rest pose, open and close the mouth once;
    \item rest pose, stretch the mouth and contract the platysma muscle once;
    \item flex, extend and rotate the neck to extreme positions, neutral expression.
    \item cervical extension pose, swallow once;
    \item cervical extension pose, open and close the mouth once;
    \item cervical extension pose, stretch mouth and contract the platysma muscle once;
    \item side-bend cervical on both sides, neutral expression;
    \item neck-shifting (twisting neck), neutral expression;
\end{itemize}
The above actions are deliberately designed for comprehensive disclosure of personalized head and neck motion space together with larynx motion. 
Similarly, we register the template mesh with recovered dynamic sequences and use the registered sequences for learning pose blendshapes and larynx deformation.

\paragraph{Larynx motion during speaking}
Besides actions like mouth stretching and swallowing, another type of action that affects larynx motion greatly is speaking. To further analyze the speaking-related larynx motion and nuanced variations, we introduce an efficient way to capture larynx and speaking-caused mouth motion simultaneously to establish their connections. Instead of relying on complicated 3D reconstructions, we analyze the larynx and mouth movements on RGB images and normal maps estimated via photometric stereo. Specifically, we efficiently track the larynx in normal maps while extracting mouth movements using off-the-shelf expression identification techniques (please see Sec.~\ref{sec:transformer} for details).

\subsection{Data Finalization}
\label{sec:data_formulation}

With ultrasound scans, aligned head and neck geometries, and physically-based textures, we can finalize our data. 
As illustrated in Fig.~\ref{fig:dataset}, our comprehensive multi-modal dataset captures the full spectrum of head and neck and consists of 624 identities with 16078 mesh registrations. From ultrasound scans and static geometry scans of the same subject, we obtain the joint positions of the head and neck skeleton at rest-pose $\mathcal{J}^r_\text{U}$ and skeleton-aligned head and neck meshes $\mathcal{N}_\text{U}^r$ from 30 identities, where the symbol $\mathcal{N}$ means meshes recovered under neutral expression, the superscript $r$ denotes rest-pose, while subscript $U$ means the ultrasound data.
To cover more varieties of identities, expressions, and poses, we aggregate samples from ICT-FaceKit with our static captures at the rest pose, and obtain $\mathcal{N}^r = \{ \mathcal{N}_\text{H}^r, \ \mathcal{N}_\text{I}^{r} \} \in \mathbb{R}^{3N\times (N_\text{H}^r+N_\text{I}^{r} )}$, with larynx removed, where subscript H means HACK data and I means ICT-FaceKit data, and specifically $N_\text{H}^r=624, N_\text{I}^{r}=600$. Notice that $\mathcal{N}_\text{U}^r$ is a subset of $\mathcal{N}_\text{H}^r$. We then subtract $\mathcal{N}_\text{H}^r$ from the original meshes without removing the larynx to obtain the pure larynx geometries $\Gamma^r_\text{H}$. Further, we have also captured subjects with various expressions in static scans, we denote this data as $\mathcal{T}_\text{H}^{r}$, where symbol $\mathcal{T}$ means meshes making expressions, $r$ denotes rest pose, and we have 208 such subjects. Notice we do not remove the larynx geometry from $\mathcal{T}$. For the captured dynamic performance sequences, we have the mesh sequence $\mathcal{D}_\text{H}=\{\mathcal{T}_\text{H}^p (i)\}$, $i$ denotes the $i$-th subject, that both making expressions and changing head and neck poses, where we have twelve identities in total. At last, we use all recovered textures, including diffuse, specularity, and normal maps, together with textures from the online dataset as $\mathcal{X}$ for our appearance learning, which consists of 360 identities. We have also captured 2D images and normal maps ${\bm \zeta}_\text{H}^{c}$, ${\bm \zeta}_\text{H}^{n}$ of the speaking sequences with a total of 6000 frames for larynx motion applications.
In summary, the full collected dataset for our HACK model learning is:
\begin{equation}
\mathbf{DATA} =  \{ \mathcal{J}^r_\text{U}, \mathcal{N}_\text{U}^r, \mathcal{N}^r, \bm{\Gamma}^r_\text{H}, \mathcal{T}_\text{H}^{r}, \mathcal{D}_\text{H}^{\vphantom{r}}, \mathcal{X}, {\bm \zeta}_\text{H}^{c}, {\bm \zeta}_\text{H}^{n} \},
\end{equation}
with skeleton $\mathcal{J}^r_\text{U}$ well matching neutral meshes in rest pose $\mathcal{N}_\text{U}^r$, combined neutral meshes $\mathcal{N}^r = \{ \mathcal{N}_\text{H}^r, \ \mathcal{N}_\text{I}^{r} \}$, larynx geometry $\Gamma^r_\text{H}$ is calculated from scanned neutral meshes $\mathcal{N}_\text{H}^r$ and dynamic sequence $\mathcal{D}_\text{H}=\{\mathcal{T}_\text{H}^p (i)\}$.
Our comprehensive multi-modal dataset covers both internal anatomical structures and external appearances, which builds an important biomechanical prior for HACK, allowing for anatomically-consistent and realistic modeling and animation.

\section{learning \modelname}
\label{sec:methodology}

Recall that our HACK model provides anatomy-consistent and expressive disentanglement of the head and neck regions through a set of parameters, i.e., $\{ \bm{\beta},\bm{\psi},\bm{\theta}, \eta, \tau, \bm{\alpha}\}$ to represent the shape, expression, skeletal pose, larynx motions, and appearance, respectively. Similarly to previous models~\cite{loper2015smpl,FLAME:SiggraphAsia2017}, we utilize blendshapes to transfer the parameters into geometric deformations. We hence formulate the mean shape $\mathbf{\bar T}$, shape blendshapes $\mathcal{S}$, expression blendshapes $\mathcal{E}_{\bm \beta}$, pose blendshapes $\mathcal{P}_{\bm \beta}$ and larynx function $L(\cdot)$, as in Eqn.~\ref{eq:T-beforetheta}. 
We further introduce learning the mapping $\mathcal{M}_{E}$ and $\mathcal{M}_{P}$ from the identity parameter ${\bm \beta}$ to expression and pose blendshapes, respectively, to augment the capability for personalized modeling. Besides, for the LBS skinning process, we construct the joint position regressor $J(\cdot)$ and skinning weights $\mathcal{W}$. 
In a nutshell, we set out to learn the model parameters including $\{ \mathbf{\bar T}, \mathcal{S}, J,\mathcal{E}_{\bm{\beta}},\mathcal{P}_{\bm{\beta}}, L,\mathcal{W} \}$ in a two-stage optimization framework based on various data modalities. 
We first learn the shape blendshapes $\mathcal{S}$, joint regressor $J$, expression and larynx blendshapes $\mathcal{E}_{\bm \beta}$, $\mathcal{L}_{\bm \beta}$ from the captured data under the rest pose (Sec.~\ref{HNSkeletonRegression}). Then, using the dynamic and sequential data, we further estimate the pose blendshapes $\mathcal{P}_{\bm \beta}$ and skinning weight $\mathcal{W}$ with carefully designed regularizations (Sec.~\ref{sec:learn_dynamic}).

\subsection{Static Geometry Modeling}
\label{HNSkeletonRegression}

\paragraph{Identity Learning}

The identity learning involves computation of the mean head and neck mesh $\mathbf{\bar T}$ and building shape blendshapes $\mathcal{S}$, where we use the static neutral meshes in rest pose $\mathcal{N}^r$ as mentioned in Sec.~\ref{sec:data_formulation}. The mean template mesh $\mathbf{\bar T}$ is calculated as the mean of $\mathcal{N}^r$.
We then apply the PCA on the displacement between $\mathcal{N}^r$ and $\mathbf{\bar T}$ to find the shape blendshapes $\mathcal{S} =[\mathbf{S}_1,\dots,\mathbf{S}_{|\bm{\beta}|}]\in \mathbb{R}^{3N\times |\bm{\beta}|}$ by only preserving the first $|\bm{\beta}|$ principal components. Notice that for $\mathcal{N}^r$, we have removed the larynx from the neutral meshes, hence the shape blendshapes $\mathcal{S}$ does not contain the larynx geometry.

\paragraph{Larynx blendshapes} Similar to the shape blendshapes, we learn the larynx blendshapes from $\bm{\Gamma}^r_\text{H}$ using the PCA decomposition. 
Instead of conducting PCA on 3D geometry directly, we model the larynx geometric displacement w.r.t. neutral mesh in the UV space for convenience. Specifically, we un-warp $\bm{\Gamma}^r_\text{H}$ into a 2D image through texture atlasing and storing the vertex displacement into corresponding uv coordinate. We conduct PCA on the obtained images and obtain larynx blendshapes basis $\mathrm{L}_i (\tau=0) \in \mathbb{R}^{3 \times H \times W}$, where $\tau=0$ means the larynx blendshapes basis is computed at the rest pose.

\paragraph{Head and neck skeleton regression}
Here we adopt the anatomically-consistent skeletal structure from Sec.~\ref{sec:HNSkeletonModel}. 
Similar to preivious models~\cite{FLAME:SiggraphAsia2017,loper2015smpl}, we learn a mapping function $J({\bm \beta}): \mathbb{R}^{|{\bm \beta}|} \mapsto \mathbb{R}^{3K}$ which takes the identity parameter ${\bm \beta}$ as input and regresses the joint positions.
We use the aligned joints and neutral mesh data, i.e., $\mathcal{J}^r_\text{U}$ and $\mathcal{N}_\text{U}^r$, to learn $J$, by minimizing the Euclidean distance loss:
\begin{equation}
    E_\text{joint}=\sum_{i}\|J({\bm \beta}_\text{U}^{\vphantom{r}} (i))-\mathcal{J}_\text{U}^r (i)\|_2^2,
\end{equation}
where ${\bm \beta}_\text{U}^{\vphantom{r}} (i)$ is the identity parameter of the $i$-th subject estimated from $\mathcal{N}_\text{U}^r$ using the shape blendshapes $\mathcal{S}$. With the learned $J$, we are able to predict joints $\mathcal{J}^r$ for all neutral meshes $\mathcal{N}^r$.

\paragraph{Person-specific expression blendshapes} From the static scans of subjects with 28 FACS expressions $\mathcal{T}^r_\text{H}$, we further construct the person-specific expression blendshapes for the captured subject with identity parameter $\bm{\beta}$, denoted as $\mathcal{E}_{\text{H}, {\bm{\beta}}}=[\mathbf{E}_{H, 1}^{\bm{\beta}},\dots,\mathbf{E}_{\text{H}, |\bm{\psi}|}^{\bm{\beta}}]\in \mathbb{R}^{3N\times |\bm{\psi}|}$, using the technique proposed by \citet{li2010example}. Then we have a set of person-specific expression blendshapes, denoted as $\{{\mathcal{E}_{\text{H}, {\bm{\beta}}}}\}=\{\mathcal{E}_{\text{H}, {\bm{\beta}} (\text{H}, 1)},\dots,\mathcal{E}_{\text{H}, {\bm{\beta}} (\text{H}, n)}\}$, for subjects in $\mathcal{T}^r_\text{H}$, where ${\bm{\beta}} (\text{H}, i)$ indicates the $i$-th subject in $\mathcal{T}^r_\text{H}$.
Different from previous face methods, such as FLAME~\cite{FLAME:SiggraphAsia2017} and ICT-FaceKit~\cite{LearningFaceCVPR2020}, that model a general expression blendshapes independent from identity information, we set out to model personalized expressions by learning a mapping network $\mathcal{M}_E$ which maps one's identity parameter ${\bm{\beta}}$ to its personalized expression blendshapes.
Specifically, we would like to train $\mathcal{M}_{E}$ on our set of person-specific expression blendshapes  $\{\mathcal{E}_{\text{H}, {\bm{\beta}}}\}$, with the following loss:
 \begin{equation}
    E_\text{exp}=\sum_{i} \big \| \mathcal{M}_{E}  \big ( {\bm{\beta}} (\text{H}, i)  \big ) - \mathcal{E}_{H, {\bm{\beta}} (\text{H}, i)}  \big \|_2.
    \label{eq:ME}
\end{equation}
Notice that the expression blendshapes, $\mathcal{E}_{\text{H}, {\bm{\beta}}}$, exist in a high dimensional space that may be difficult for the mapping network to learn. To address this challenge, we first apply PCA to the set of person-specific expression blendshapes, $\{{\mathcal{E}_{\text{H}, {\bm{\beta}}}}\}$, and the mapping network, $\mathcal{M}_E$, is then trained to predict the PCA weights. This allows for more efficient learning and improves the robustness while preserving personalized expressions as the predicted blendshapes rely on ${\bm{\beta}}$.
In the implementation, $\mathcal{M}_{E}$ consists of three linear layers with $64$ neurons and ReLU activation, and predicts the weights of the first 50 principal components. The training is conducted using Pytorch Adam optimizer with a learning rate of 0.0001, and converges in 2 hours on a single Nvidia Titan GPU.

\begin{figure}[t]
    \centering
            \includegraphics[width=1\linewidth]{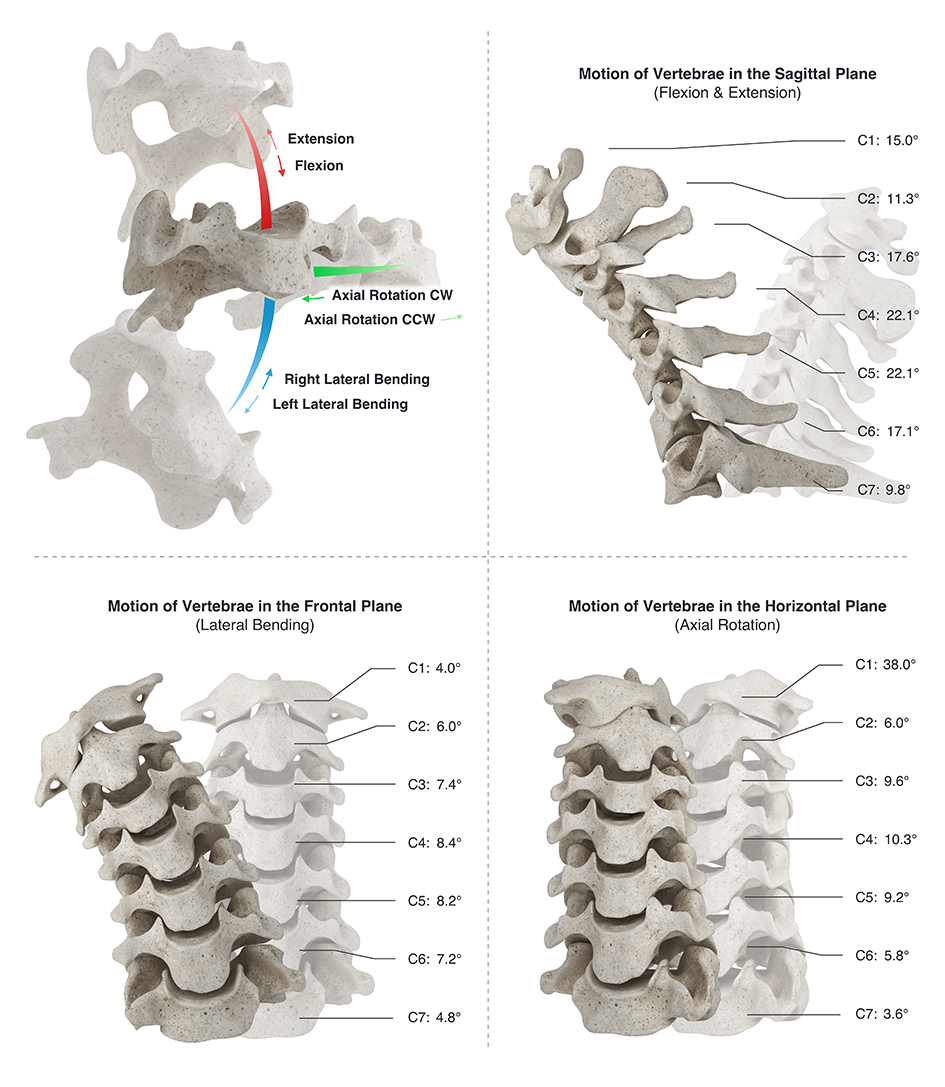}
    \caption{
    Rotation limit of the cervical spine.
    We first illustrate the degree of freedom of rotation of cervical vertebrae. Then we show the average rotation limit in angles of each cervical vertebra under different cervical motions, including flexion and extension (upper right), lateral bending (bottom left), and axial rotation (bottom right). 
    We design the rotation limit upon a comprehensive anatomical survey~\cite{anatomystandard}.
    }
    \label{fig:anatomy}
\end{figure}

\subsection{Dynamic Deformation Learning}
\label{sec:learn_dynamic}

We have obtained the shape blendshapes $\mathcal{S}$, person-specific expression blendshapes $\mathcal{E}_{\bm \beta}$, and larynx blendshapes $\mathcal{L}$ for all identities. To further model the deformations when changing the head and neck pose, we rely on the dynamic sequences $\mathcal{D}_\text{H}$ for learning the pose blendshapes, larynx function $L({\bm \beta}, \eta, \tau; \mathcal{L})$ and skinning weights $\mathcal{W}$. Notice that $\mathcal{W}$ is shared across all identities, and we hire professional animation artists to create initial $\mathcal{\tilde W}$ as a prior skinning weight.

To this end, we jointly learn $\mathcal{P}_{\bm \beta}$, $L({\bm \beta}, \eta, \tau; \mathcal{L})$ and $\mathcal{W}$ via minimizing the reconstruction loss on $\mathcal{D}_\text{H}$ as:
\begin{equation}
    E_\text{rec} = \sum_{i=1}^{ | \mathcal{D}_\text{H}^{\vphantom{p}} | }
    \sum_{j=1}^{ | \mathcal{T}_\text{H}^p (i) | } 
    \big \| \mathbf{G} \big (\bm{\beta}(i), {\bm{\psi}}(i, j), {\bm{\theta}}(i, j), \eta (i), \tau  (i, j) \big ) - \mathcal{T}_\text{H}^p (i, j) {\big \|}_2^2,
\end{equation}
where $\mathbf{G}(\cdot)$ is the geometry modeling as described in Sec.~\ref{sec:ModelFormulation}. $i$ is the index of the subject and $j$ indicates the frame in a sequence of a certain identity. Notice that $\mathcal{W}$ is updated in the skinning step as in Eqn.~\ref{eq:LBS}. Moreover, we also fine-tune the person-specific expression blendshapes $\mathcal{E}_{\bm{\beta}}$ during the training process for better modeling the neck motion in this stage.

Besides supervision from captured meshes, we additionally design several regularization terms, including joint rotation limits, pose regularization, and collision penalization, based on anatomy and physical priors to improve the robustness of training. 

\paragraph{Joint rotation limits}
The moving range of the human cervical vertebrae is limited due to its unique anatomy structure, as illustrated in Fig.~\ref{fig:anatomy}. 
To simulate the head and neck motion correctly and obtain realistic model deformations, we limit the rotation angles that can be applied to each joint according to Fig.~\ref{fig:anatomy}. Specifically, we convert the pose parameter $\bm{\theta}$ to Euler angles and enforce penalization when any angle of a certain joint exceeds the corresponding limits. The penalization is formulated as the following loss:
\begin{equation}
    E_\text{rot}=\sum_{i, j} \sum_{1\le k \le K} \big \| \max\{|\theta^k (i, j)| - b^k,0\} \big \|_1,
\end{equation}
where $i, j$ denote the identity and frame in dynamic sequence respectively. $\theta^k$ is the Euler angle of $k$-th joint of given pose $\bm{\theta}$, $b_k$ is the corresponding rotation limits exhibited as in Fig.~\ref{fig:anatomy}.
\paragraph{Adjacent joints consistency} Further, the joints do not rotate independently, and the rotations of adjacent cervical vertebrae exhibit certain similarities. Hence, we minimize the rotation differences of Euler angles between the adjacent joints, as:
\begin{equation}
    E_\text{sim}=\sum_{1\le i < K}\|\theta_i-\theta_{i+1}\|_2^2,
\end{equation}
where $\theta_i\in \bm{\theta}$ is the $i$th rotation vector of given pose $\bm{\theta}$.
\paragraph{Collision penalization}
We find that the neck modeling is prone to noise and inaccuracies due to heavy occlusion, particularly when the head is tilted. To address this, we enforce collision penalization to ensure that the cervical spine is not intersected with the neck skin, and we re-use the collision term $E_{col}$ proposed in~\cite{Hasson_2019_CVPR}.

\paragraph{Temporal smoothness}
As the per-frame learning scheme leads to noises, we introduce temporal consistency terms for parameters related to motion, including $\bm{\psi}, \bm{\theta}, \eta, \tau$. 
For every scalar element $v$ in $\{ \bm{\psi}, \bm{\theta}, \eta, \tau \}$, we aggregate the estimated $v$ for a captured sequence $s$ and form the vectors $\{ v(t) \}_s$ for each $v$, where $t$ indicates the frame.
We enforce the Lipschitz continuity and minimize the second derivative of $\{ v \}_s$ as follows:
\begin{equation}
    E_\text{tem}=\sum_s \lambda_v \sum_{ v(t) \in \{ v(t) \}_s } (\lambda_1 \max\{|\frac{d [v(t)]}{dt}|-\epsilon,0\}^2+\lambda_2(\frac{d^2 [v(t)]}{dt^2})^2),
\end{equation}
where $\lambda_v,\lambda_1, \lambda_2$ are the weighting factors, $\epsilon$ is threshold that to tolerate small deviations. For $\bm{\psi}$, $\lambda_v=1,\lambda_1=1, \lambda_2=0.01, \epsilon=0.1$; for $\bm{\theta}$, $\lambda_v=1,\lambda_1=1, \lambda_2=5000, \epsilon=0.15$; for $\bm{\eta}$, $\lambda_v=1,\lambda_1=1, \lambda_2=10, \epsilon=0.005$; for $\bm{\tau}$, $\lambda_v=1,\lambda_1=1, \lambda_2=1, \epsilon=0.001$.

\paragraph{Regularization}
Similar to the temporal smoothness of parameters, we want to ensure the generated geometries are also smooth in temporal dimensions. 
We apply the Laplacian smoothness penalty $E_\text{smo}$ for all pose blendshapes. Also, $E_\text{ski}=\|\mathcal{W}-\mathcal{\tilde W}\|_2^2$ is introduced to regularize skinning weight to not deviate from artist-created initial weights too much.
\vspace{0.5cm}
The final optimization loss is the combination as follows:
\begin{equation}
    E=E_\text{rec}+E_\text{rot}+E_\text{sim}+E_\text{col}+E_\text{tem}+E_\text{smo}+E_\text{ski},
\end{equation}
where the weighting factors for each penalty are 1e5, 1e6, 5e3, 5e5, 1e6, 5e-2, 1, respectively.
During training, we estimate the person-specific pose blendshapes $\mathcal{P}_{\bm \beta}$ with pose parameters $\bm{\theta}$, expression parameters $\bm{\psi}$ and larynx parameters $\eta,\tau$, jointly.
During the learning process, \modelname can be considered as a differentiable layer, and the optimization process is carried out using Adam optimizer in PyTorch. Each dynamic performance sequence has an average length of 850 frames, and the overall loss converges after 45000 epochs with the learning rate set at 0.001.

After every person-specific pose blendshapes converge for each identity, we have a set of person-specific pose blendshapes, denoted as $\{{\mathcal{P}_{{\bm{\beta}}}}\}=\{\mathcal{P}_{{\bm{\beta}} (\text{H}, 1)},\dots,\mathcal{P}_{{\bm{\beta}} (\text{H}, n)}\}$, where ${\bm{\beta}} (\text{H}, j)$ indicates the $j$-th subject in $\mathcal{D}_\text{H}$.
Still, instead of modeling a general pose blendshapes as previous methods~\cite{FLAME:SiggraphAsia2017,loper2015smpl}, we further model personalized pose blendshapes by learning the mapping network $\mathcal{M}_{P}$ that maps identity code to its personalized pose blendshapes.
The learning process of $\mathcal{M}_{P}$ is the same as expression mapping network $\mathcal{M}_{E}$ as in Eqn.~\ref{eq:ME}. Similarly, to reduce the dimensionality of this even larger space $\mathcal{M}_{P}\in \mathbb{R}^{3N\times 72}$, PCA is applied on $\{\mathcal{P}_{\bm{\beta}}\}$ similar to learning $\mathcal{M}_{E}$.Due to the complexity of capturing dynamic head and neck performance sequences, we captured twelve identities that have person-specific pose blendshapes for $\mathcal{M}_{P}$ to learn. However, modeling personalized pose blendshapes instead of the general ones is still proved effective that better restores the identity-dependent characteristics, which will be evaluated in Sec.~\ref{sec:ablation}. As more dynamic head and neck performance sequences are captured, modeling personalized motion features, including expression and pose blendshapes, will demonstrate its strong capability and continue to benefit the community of parametric models.

\subsection{Appearance Learning}
For authentic appearance generation, we create a
parametric appearance model from our physically-based texture dataset, including diffuse albedo, specular intensity, and normal map. Using the multi-view photometric system, we 
capture the rich physically-based appearances for the whole head-and-neck regions of various subjects, and subsequently unwarp the appearance into unified texture maps from the neutral scans at the rest pose. We follow previous work~\cite{HTML,Sculptor} to unify all the physically-based texture data, and subsequently perform principal component analysis (PCA) using singular value decomposition directly to obtain the principal components in the UV space. To this end, we obtain the appearance model $\mathbf{A}(\bm{\alpha})$ to generate realistic textures from a random appearance parameter $\bm{\alpha}$.
Since our textures have uniform texture UV mapping as our template skin mesh, we could directly apply generated physically-based textures with various shapes and produce a photo-realistic appearance.

\section{Applications}
\label{sec:application}

As a parametric model that tackles the head and neck as a whole with expressive controls, HACK is differentiable and compatible with standard CG software, making it suitable for a range of applications like geometric fitting, animation, and inference. HACK also enables fine-grained analysis of the inter-correlation of head and neck. In the following, we demonstrate various applications like motion synthesis or transfer towards such unique characteristics.

\begin{figure}[t]
    \centering
        \includegraphics[width=\linewidth]{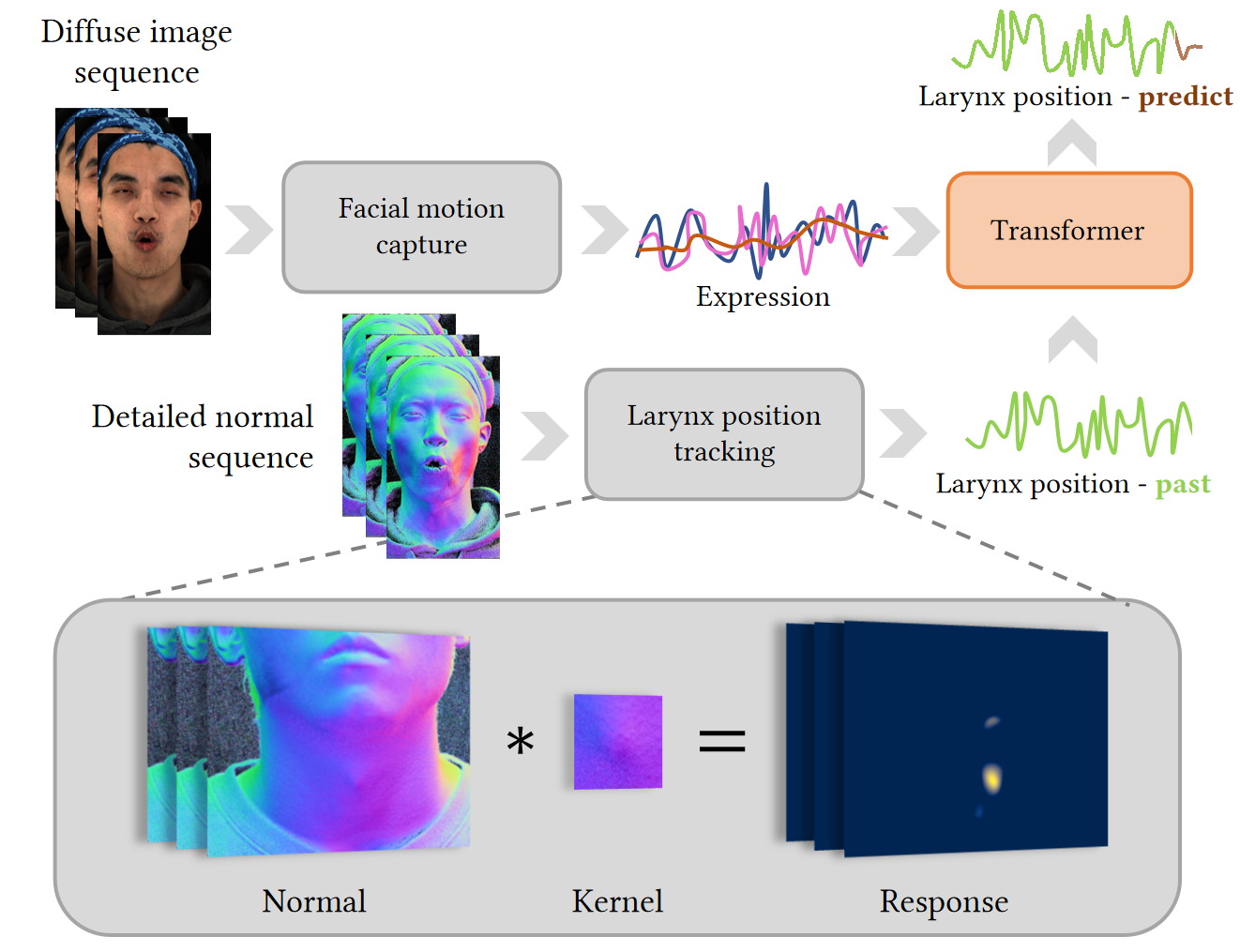}
    \caption{The pipeline of generating larynx motion sequence using transformer. After extracting expressions and larynx positions from captured sequences, we train the transformer to predict larynx motion in an auto-regressive manner.}
    \label{fig:transformer}
\end{figure}

\begin{figure*}
    \centering
        \includegraphics[width=\linewidth]{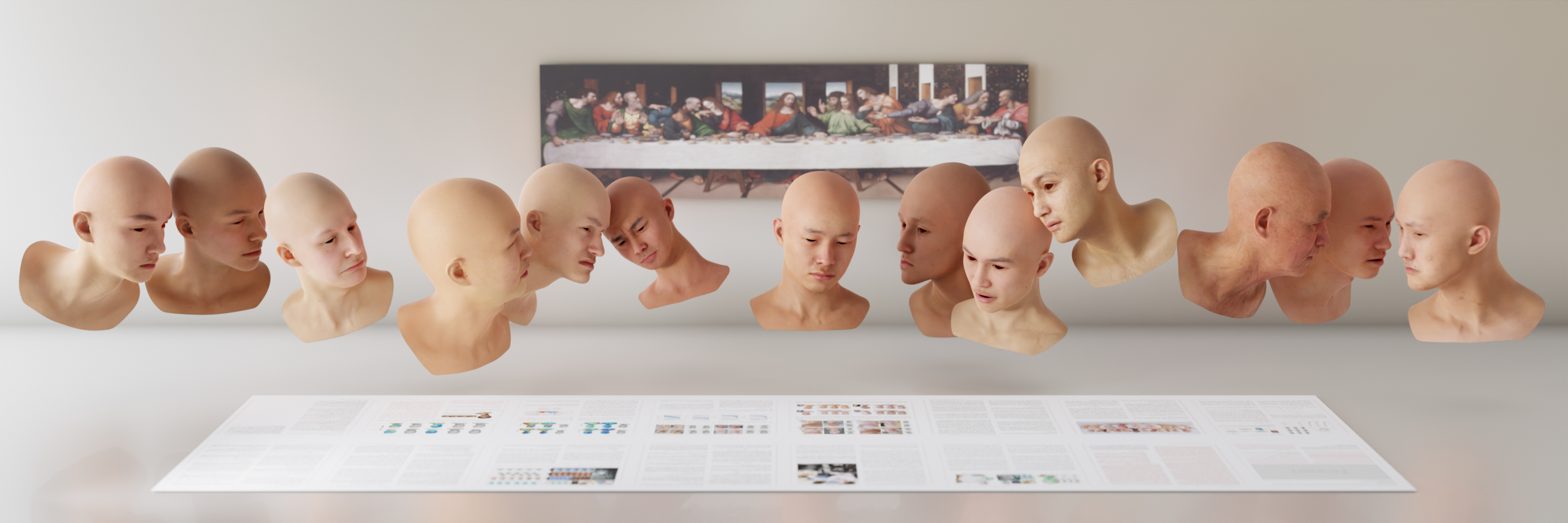}
    \centering
    \caption{
Samples using \modelname, of various identities, head and neck poses, expressions, and appearance. 
Our model demonstrates its strong ability for realistic modeling, animation and rendering.
    }    \label{fig:generation}
\end{figure*}

\subsection{Motion Synthesis from Head to Neck}
\label{sec:transformer}

The head poses and facial expressions are highly correlated with the skeletal pose of the cervical spine and neck deformation. Take the speaking action as a typical example, where the movement of the larynx is coordinated with the movements of the mouth, which could be faithfully captured by the expression parameters, as it controls the vocal cords.
Hence, we use the transformer architecture from FaceFormer~\cite{faceformer2022} to regress the larynx movement sequence $\{ \tau \}_m$ from $\bm{\Psi}$, as shown in Fig.~\ref{fig:transformer}. 

We train the transformer using our captured larynx motion sequences ${\bm \zeta}_\text{H}^c$ and ${\bm \zeta}_\text{H}^n$, where ${\bm \zeta}_\text{H}^c$ and ${\bm \zeta}_\text{H}^n$ are terms of RGB images and normal maps, respectively. 
We utilize the facial tracker NPFA~\cite{Zhang2022VideoDriven} to predict the expression parameters from ${\bm \zeta}_\text{H}^c$.
We also adopt a novel convolutional method to track the movement of the larynx in the vertical direction directly from image inputs. 
Specifically, we define a kernel $\mathbf{k}$ of size $70\times 70\times 3$ denoting the generic shape (represented as normal) of the larynx.
We extract the larynx position by finding the maximum response on the dot product of $\mathbf{k}$ and ${\bm \zeta}_\text{H}^n$ in a convolutional manner, which can be formulated as follows:
\begin{equation}
    \bar{\tau}(i, j) \leftarrow \arg\max_{x} \big [ \mathbf{k} * {\bm \zeta}_\text{H}^n (i, j) \big ] (x) - \tau_0 (i),
\end{equation}
where $*$ denotes the convolution operator; $x$ denotes the spatial position in normal maps; $\tau_0 (i)$ is the initial position of the larynx in the rest pose of the $i$-th identity; $j$ denotes the frame index in the sequence.
We follow the framework of the FaceFormer. In particular, we fed the expression $\bm{\Psi}$ into the attention layers of our transformer, and use the past sequence $\bar{\tau}(i, 1\dots k-1)$ to predict the next signal $\bar{\tau}(i, k)$. Once trained, the adopted transformer is able to predict the larynx position sequence $\{ \tau \}$ from unseen expressions $\{ \bm{\Psi} \}$ for novel speaking. Then, we can use $\{ \tau \}$ and $\{ \bm{\Psi} \}$ to synthesize the coordinated larynx motion and speaking action. 

Similarly, we can capture the correlations between the head pose and the skeletal pose of the cervical spine to effectively drive a HACK model with existing facial motion capture techniques, like DECA~\cite{DECA:Siggraph2021} and Apple ARKit~\cite{arkit}.
Specifically, we calculate the facial orientation using the global rotation of the head skull (the o-c1 joint) on our dynamic performance sequences $\mathcal{D}_\text{H}$, and set up a small MLP that consists of 2 layers with 512 neurons to learn the mapping from facial orientations to pose parameters $\bm{\theta}$ in our definition.
Then, with the correct mapping, we can manipulate the HACK model to display a range of personalized expressions, poses, and appearances with accurate neck deformations.

\subsection{Cross Species Motion Transfer}
\label{sec:transfer}

Motion transfer or retargeting is the process of transferring expressions, poses, and other motions from one character to another. In this section, we demonstrate another unique ability of HACK: the ability to transfer expressions and poses from humans to another mammal model, in this case, a giraffe. Existing generic models cannot perform this type of transfer as they do not model neck motion. However, with HACK's neck modeling ability, the transferred neck motion achieves a high level of realism due to the neck skeleton's structural similarity between the characters.

We start by using a giraffe model with defined neck skeleton joints, which has been created by artists. We require the giraffe model to have the same topology as the HACK template model and in rest pose. Next, we apply the estimated expression and pose parameters, which were fitted using human models, to the giraffe model. Because HACK's head and neck skeleton is anatomically consistent and mammals share the same neck skeleton structure, the transferred poses on the giraffe are highly realistic, as demonstrated in Fig.~\ref{fig:giraffe}.

\section{Results}
\label{sec:experiment}

\subsection{Model Evaluation}

\paragraph{Qualitative results}

\begin{figure*}
    \centering
    \begin{overpic}[width=\linewidth]{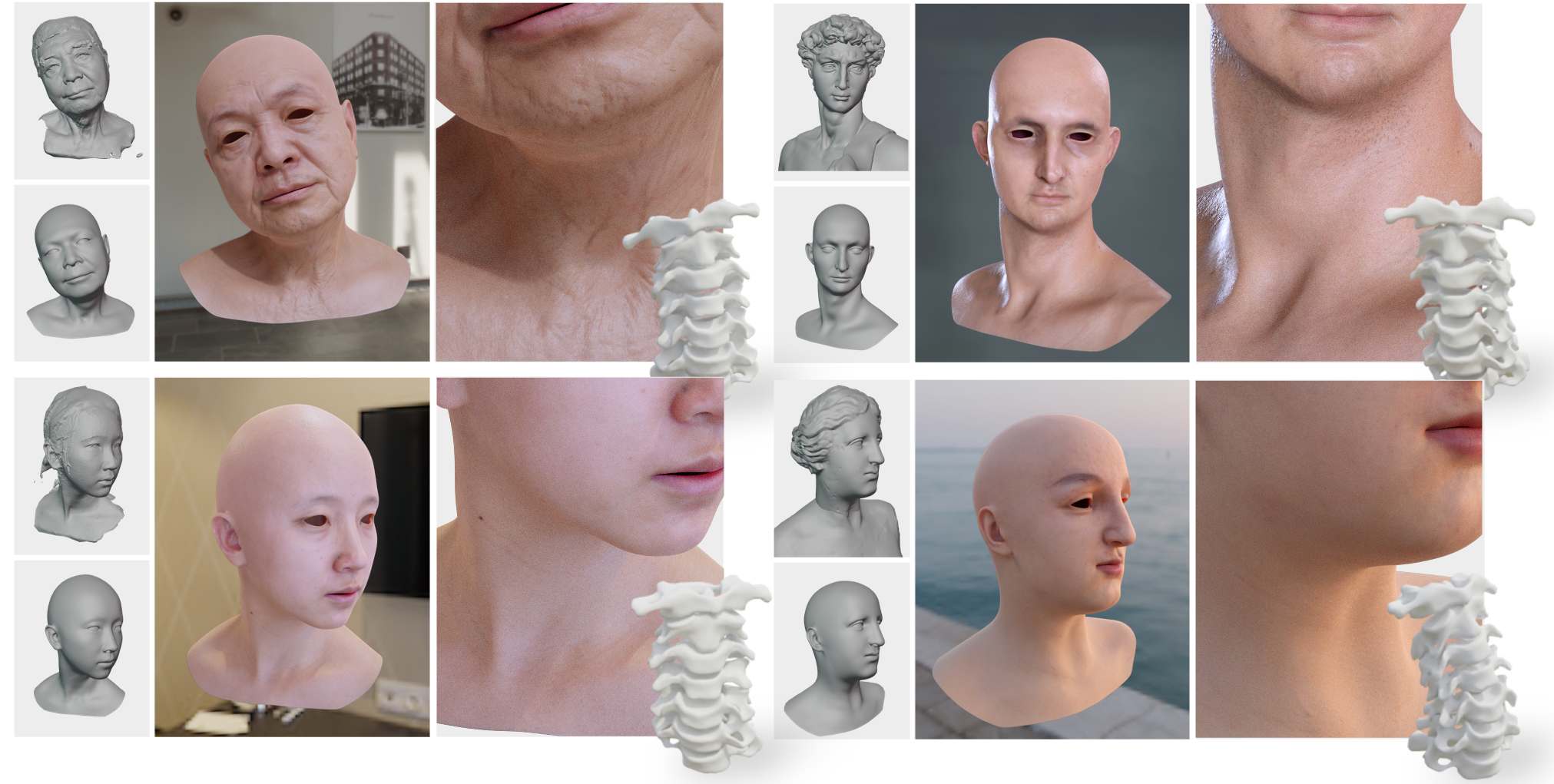}
    \put(4,1){ \color{black}{(a)}}
    \put(17,1){ \color{black}{(b)}}
    \put(36,1){ \color{black}{(c)}}
    \put(53,1){ \color{black}{(a)}}
    \put(67,1){ \color{black}{(b)}}
    \put(85,1){ \color{black}{(c)}}
    \end{overpic}
    \vspace{-0.7cm}
    \centering
    \caption{
Registration results of \modelname on our testing dataset and masterpiece statues. 
As discussed in Sec.~\ref{sec:application}, we can fit \modelname on different data, ranging from our captured unseen scans to masterpiece statues. In the left part, we first demonstrate the registration of an old man and a young lady from our captured data, then, in the right part, we show the registration of \textit{David}, a masterpiece of Renaissance sculpture, and \textit{Venus de Milo}, an ancient Greek sculpture that was created during the Hellenistic period, both with an elegant neck.
For each case, we show (a) the target and \modelname's registration; (b) the registration with appearance sampled from \modelname's appearance space; (c) the zoom-in view to demonstrate the details and corresponding neck pose.
\modelname successfully models across identities, poses, and expressions, with details restored and can be realistically rendered with appearance applied.
    }
    \label{fig:fitting}
\end{figure*}

\begin{figure*}
    \centering
    \includegraphics[width=\linewidth]{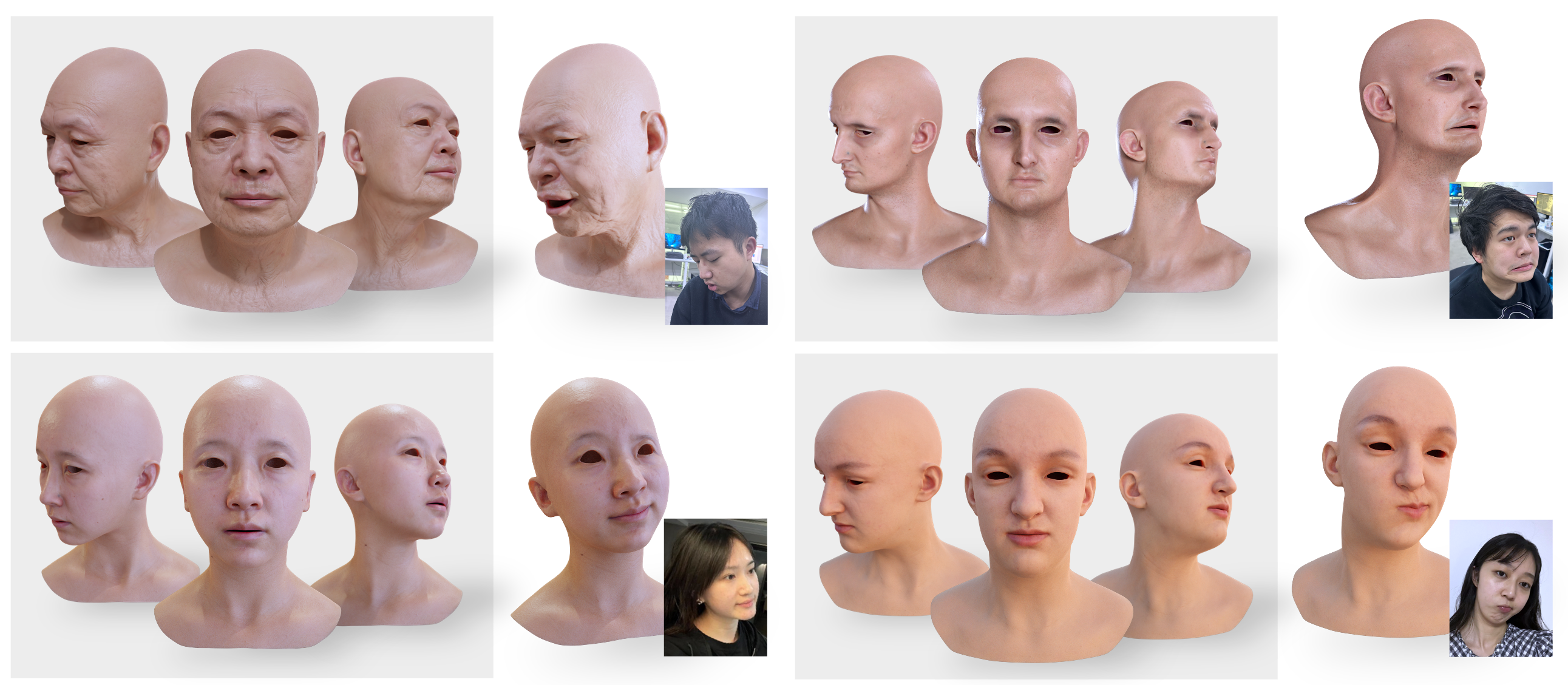}
    \vspace{-0.3cm}
        \begin{tabular}{>{\centering\arraybackslash}p{5.4cm}>{\centering\arraybackslash}p{2.8cm}>{\centering\arraybackslash}p{5.4cm}>{\centering\arraybackslash}p{2.7cm}}
    (a) & (b) & (a) & (b)
    \end{tabular}
    \centering
        \caption{
Novel pose and inference results of previous \modelname registrations.
For each case, we provide (a) the same identities with synthetic novel poses; (b) driven results using facial performance capture, as discussed in Sec.~\ref{sec:application}.
Note that for \textit{David}, our driving result provides the realistic bowstring effect where the platysma is contracting, thanks to our joint modeling of the head and neck.
As shown in the figure, \modelname provides realistic driven and rendering results.
    }    \label{fig:inference}
\end{figure*}

\begin{figure*}
    \centering
    \includegraphics[width=\linewidth]{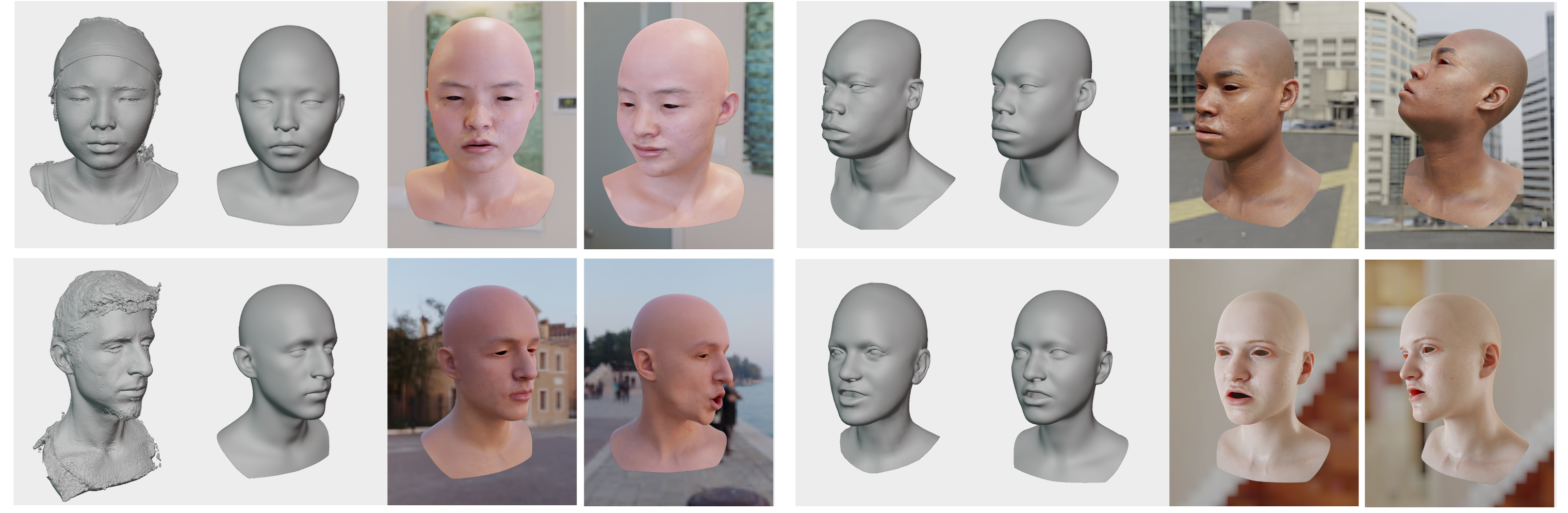}
        \begin{tabular}{>{\centering\arraybackslash}p{1.9cm}>{\centering\arraybackslash}p{1.8cm}>{\centering\arraybackslash}p{1.9cm}>{\centering\arraybackslash}p{1.9cm}>{\centering\arraybackslash}p{1.9cm}>{\centering\arraybackslash}p{1.9cm}>{\centering\arraybackslash}p{1.8cm}>
    {\centering\arraybackslash}p{1.8cm}}
    (a) & (b) & (c) & (d) & (a) & (b) & (c) & (d)
    \end{tabular}
    \centering
    \caption{
Registration results on various released dataset, including samples from FaceScape~\cite{yang2020facescape} (upper left), 3D Scan Store~\cite{3dscanstore} (upper right), MultiFace~\cite{wuu2022multiface} (lower left), and VOCA~\cite{VOCA2019} (lower right). For each sample, we show (a) the original mesh, (b) \modelname registration result, (c) the realistic rendering result using our physically-based appearance, and (d) with novel expressions and poses.
    }
    \label{fig:registration}
\end{figure*}

HACK offers personalized and anatomically-consistent controls for the neck regions by leveraging comprehensive inner biomechanical priors.
As demonstrated in Fig.~\ref{fig:generation}, by controlling the parameters of HACK, HACK can generate diverse results and even form a delicate artwork.
Being a parametric model, HACK has the ability to fit on existing meshes under various neck poses. Enhanced by modeling personalized characteristics such as expression and pose blendshapes, HACK is powerful in restoring their unique traits and nuanced personalities, as illustrated in Fig.~\ref{fig:fitting}.
HACK combines inner anatomical structures and physically-based appearance to model the full spectrum of neck motions and detailed facial expressions. It supports personalized and anatomically-consistent controls using existing facial trackers, while providing realistic animation and rendering results, as shown in Fig.~\ref{fig:inference}.
As shown in Fig.~\ref{fig:registration}, HACK demonstrates strong generalization ability by successfully registering various released datasets and generating novel appearances and poses with realistic rendering.

\begin{figure*}
    \centering
    \setlength{\tabcolsep}{0pt}
    \begin{tabular}{cccc}
        \includegraphics[width=0.25\linewidth]{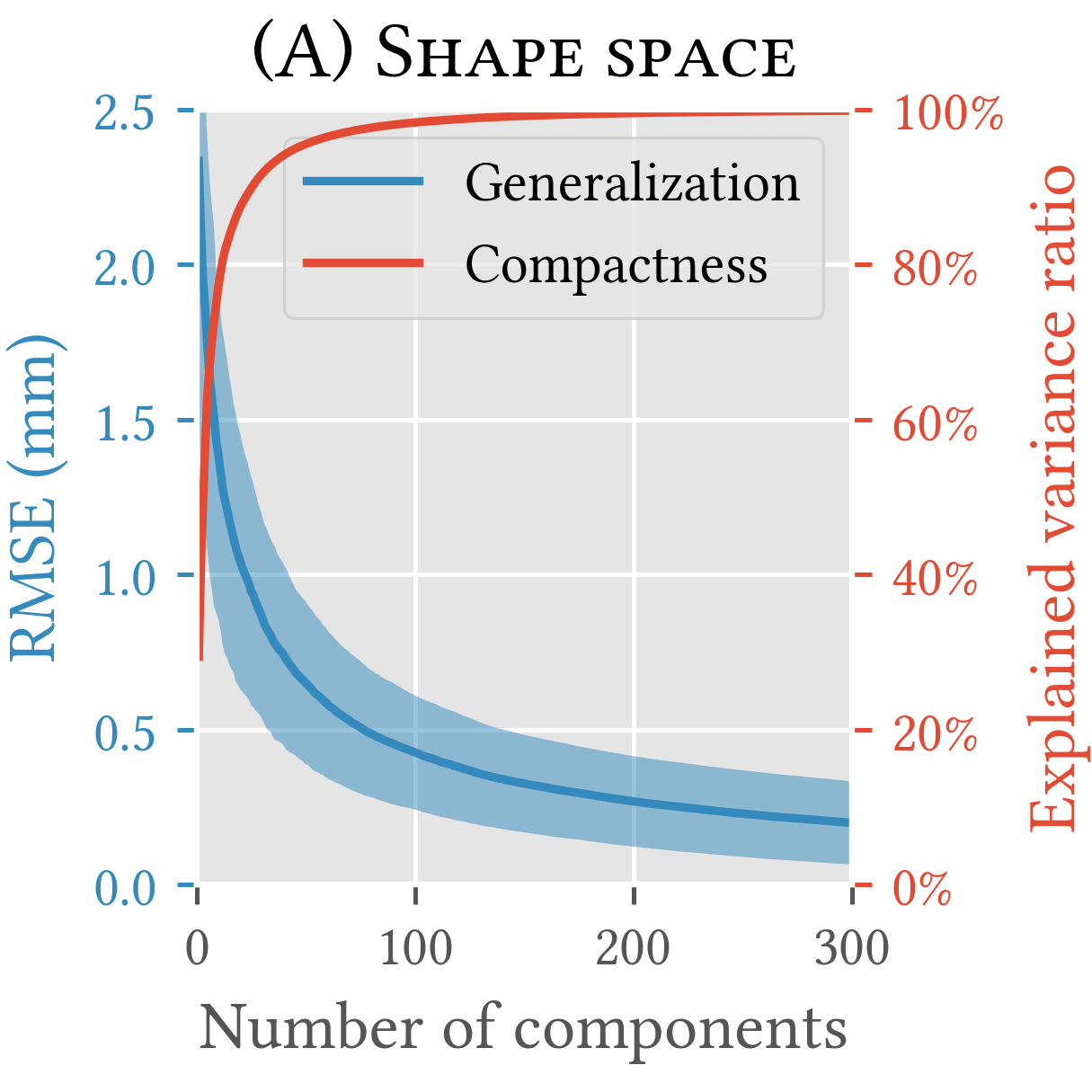} & 
        \includegraphics[width=0.25\linewidth]{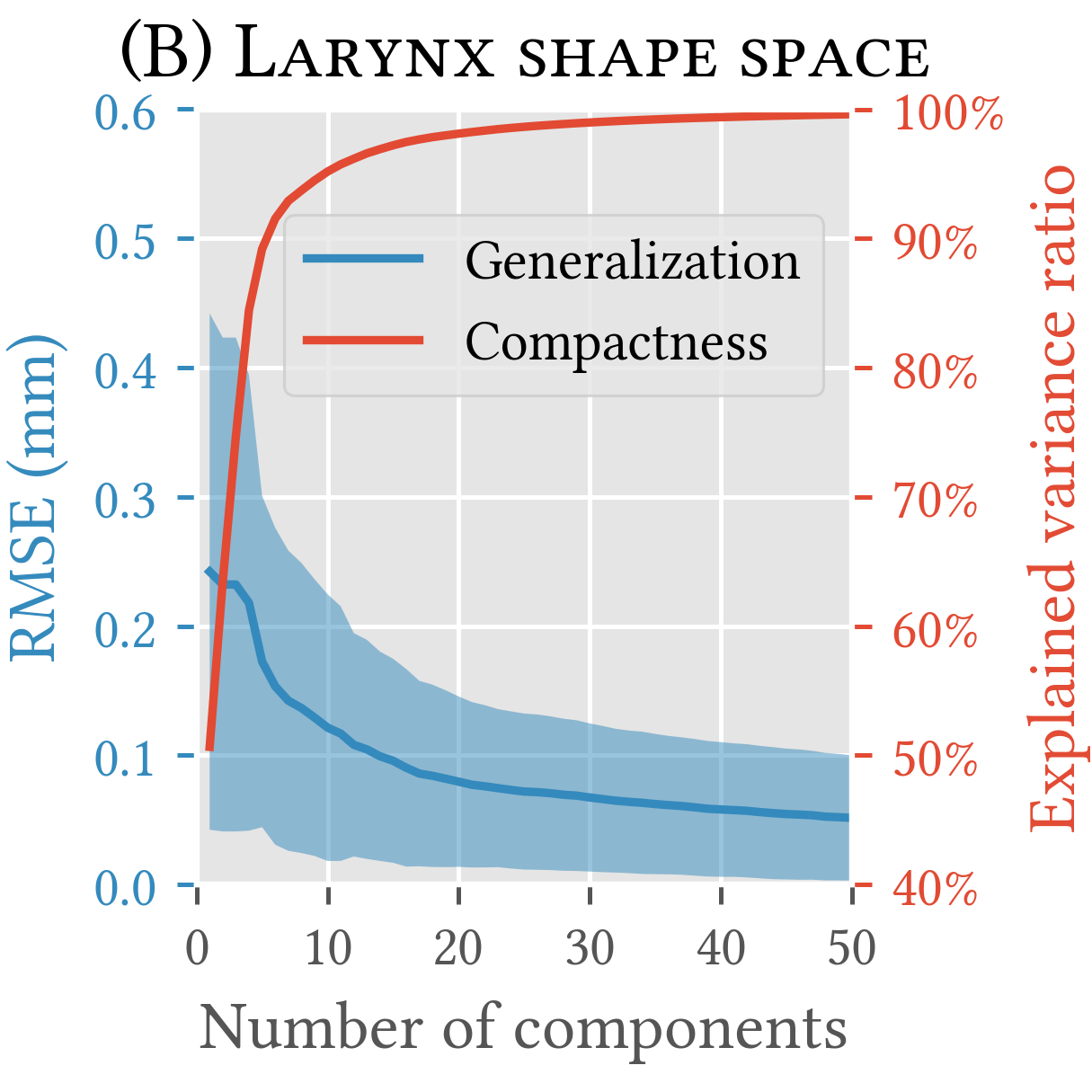} &
        \includegraphics[width=0.25\linewidth]{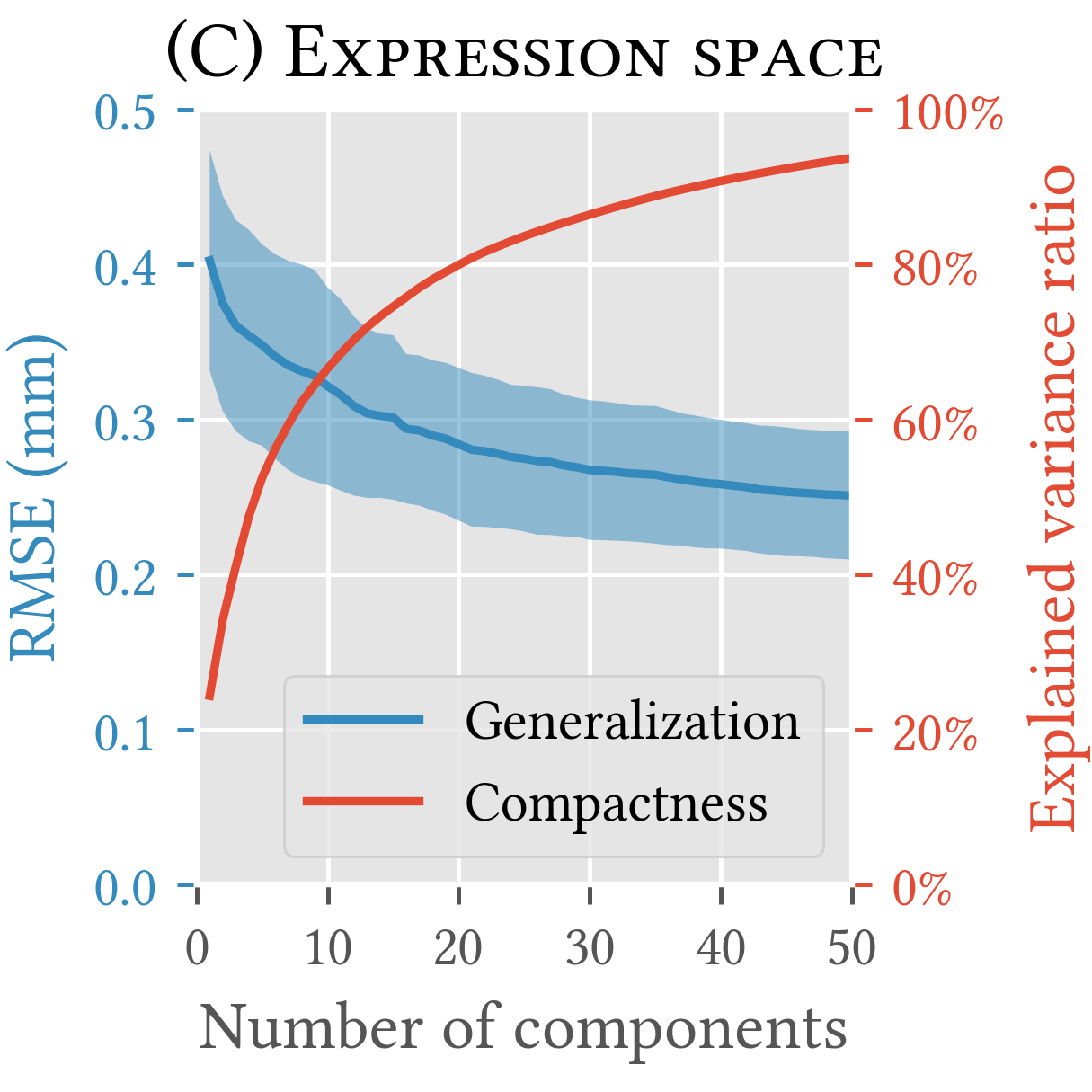} &
        \includegraphics[width=0.25\linewidth]{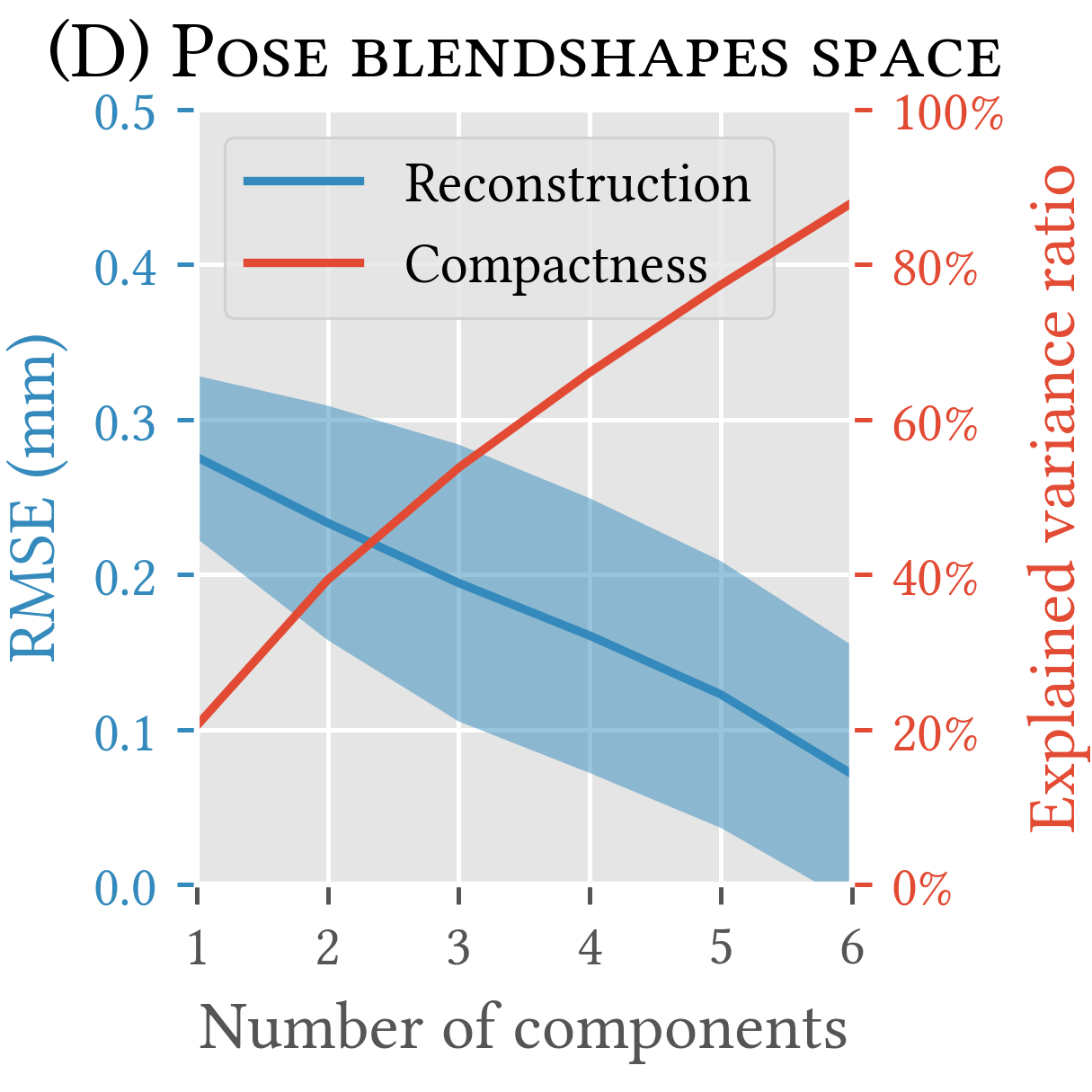}
        \\
            \end{tabular}
                                    \caption{Quantitative evaluation of compactness and generalization.}
    \label{fig:quantitative}
\end{figure*}

\paragraph{Quantitative evaluation}

Fig.~\ref{fig:quantitative} shows the compactness of learned spaces in \modelname. These red curves describe the explained-variance ratio in the training data with respect to the number of components.
In shape space, the curve in Fig.~\ref{fig:quantitative}(a) implies that with the first 50 principal components, the shape space is able to cover $95.5\%$ of the entire space. Meanwhile, 200 principal components are sufficient to express $99.5\%$ of the entire space. 
In larynx shape space, Fig.~\ref{fig:quantitative}(b) shows that the first 10 principal components achieve over $95.2\%$ of the larynx shape space, while 30 components are able to cover $99.0\%$ of the larynx shape space.
In expression blendshapes space, Fig.~\ref{fig:quantitative}(c) shows that the first 30 principal components achieve over $86.4\%$ of the expression blendshapes space, while 50 components are able to cover $93.7\%$ of the whole space.
In pose blendshapes space, Fig.~\ref{fig:quantitative}(d) shows that 6 components are able to cover $87.9\%$ of the space.

\begin{table}[t]

    \caption{
Quantitative comparison on neutral registration on the FaceScape dataset~\cite{yang2020facescape}, VOCASET~\cite{VOCA2019},  ICT-3DRFE~\cite{stratou2011effect} and Multiface~\cite{wuu2022multiface}.
    }

    \begin{tabular}{cccc}\hline\hline
       Dataset  &  FLAME & ICT-FaceKit & \modelname \\\hline\hline
       FaceScape  &  $2.401\pm 0.915$ &  $2.194\pm 1.644$ &  $\mathbf{1.929\pm 0.709}$   \\\hline
       VOCASET  &  $1.406\pm 0.280$ &  $0.958\pm 0.148$ &  $\mathbf{0.913\pm 0.133}$   \\\hline
       ICT-3DRFE  & $0.397\pm 0.045$& $\mathbf{0.366\pm 0.041}$& $0.376\pm 0.048$ \\\hline
       Multiface  & $0.943\pm 0.082$& $0.901\pm 0.120$& $\mathbf{0.842\pm 0.127}$ \\\hline
       avg. &  $1.298\pm 0.972$& $1.132\pm 1.173$& $\mathbf{1.035\pm 0.750}$ \\\hline
       \hline
    \end{tabular}

    \label{tab:comparison}
\end{table}

\begin{figure}[t]
    \centering
    \newcommand{\width}{0.25\linewidth}
    \newcommand{\widthh}{0.3\linewidth}
    \setlength{\tabcolsep}{1pt}
    \begin{tabular}{ccccc}
        \includegraphics[width=\width]{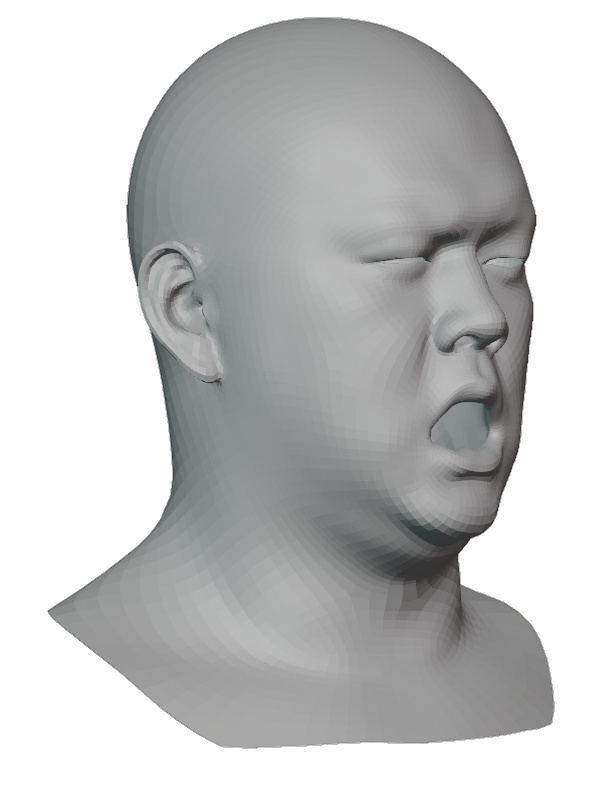} & 
        \includegraphics[width=\width]{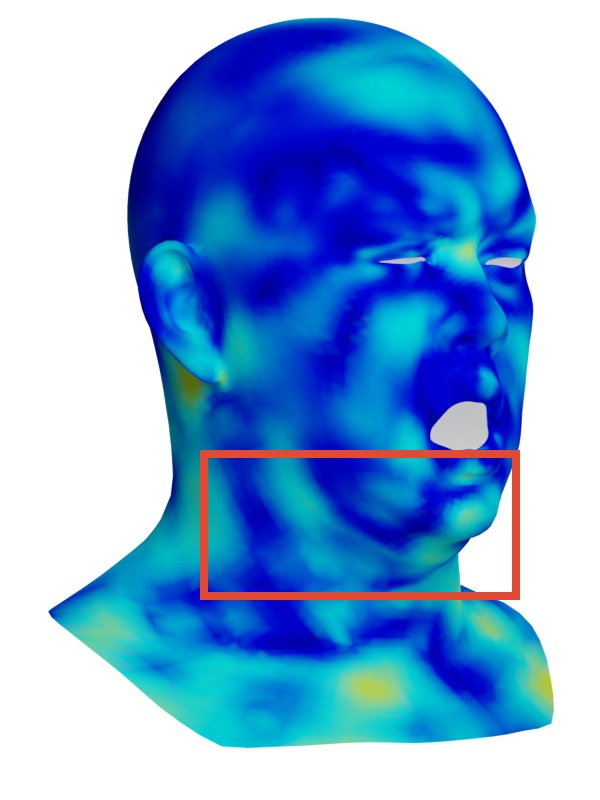} &
        \includegraphics[width=\width]{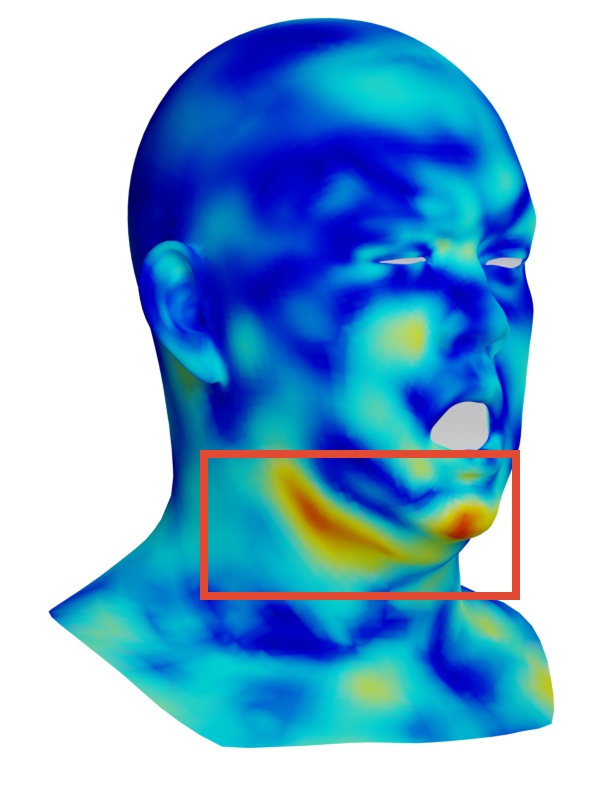} &
        \raisebox{-0cm}{\includegraphics[width=1cm]{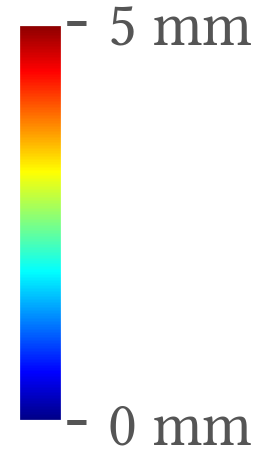}}
        \\[-0.6cm]
         & 
        \includegraphics[width=\widthh]{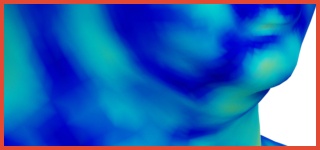} &
        \includegraphics[width=\widthh]{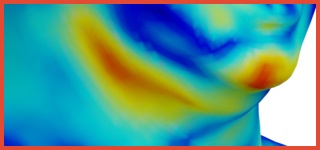}
        \\
        \includegraphics[width=\width]{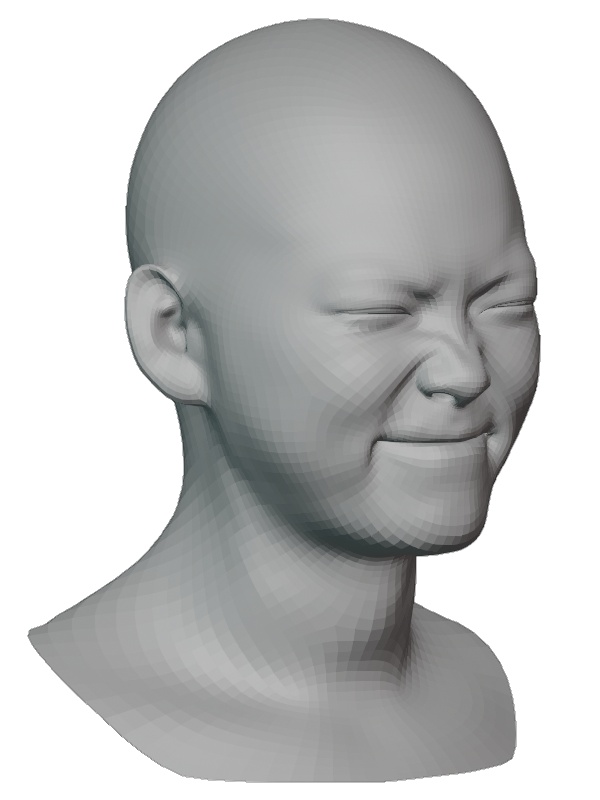} & 
        \includegraphics[width=\width]{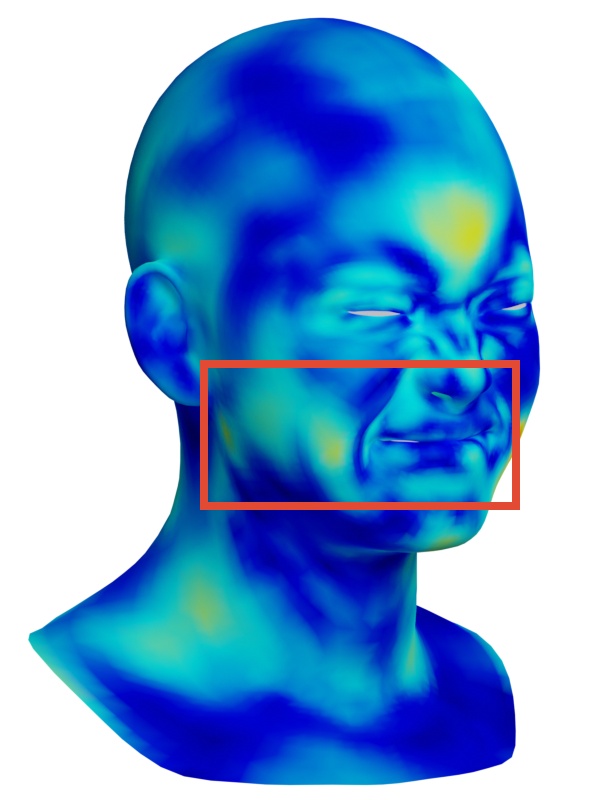} &
        \includegraphics[width=\width]{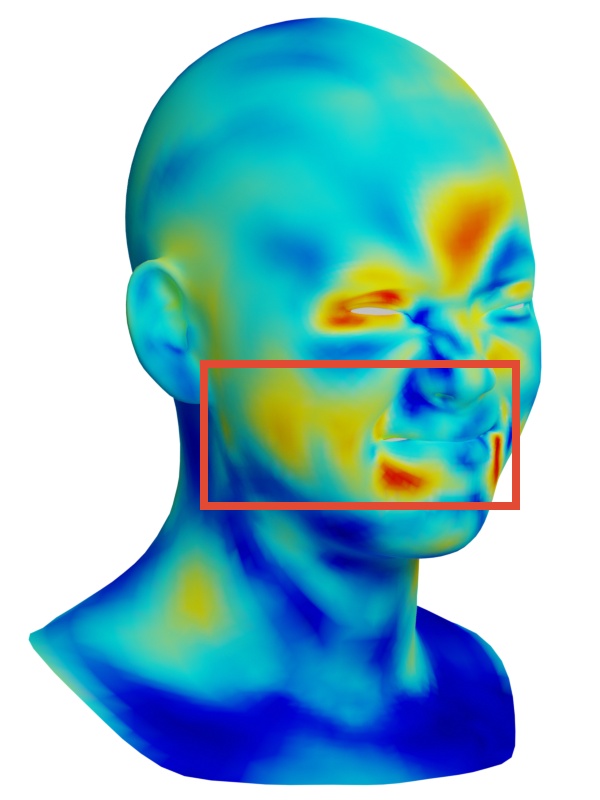} &
        \raisebox{-0cm}{\includegraphics[width=1cm]{fig/realfig/colorbar.png}}
        \\[-0.6cm]
         & 
        \includegraphics[width=\widthh]{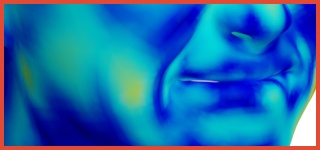} &
        \includegraphics[width=\widthh]{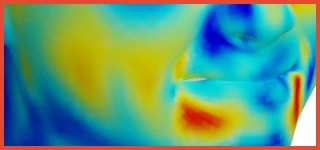}
        \\
        Target & \textbf{personalized $\mathcal{E}$} & \textbf{generic $\mathcal{E}$}
    \end{tabular}
    \caption{
    Qualitative evaluation of whether to use person-specific expressions.
        The upper performer opens his mouth but with slightly different jaw movements than normal people due to his high body fat rate. The lower performer makes a "fully compressed" face that is extremely difficult to fit expression using blendshapes. With modeling personalized expressions, our model achieves better reconstruction accuracy on rich expressions, especially around the mouth and jaw.
        }
    \label{fig:ablation_E}
\end{figure}

\begin{figure}[t]
    \centering
    \newcommand{\width}{0.25\linewidth}
    \newcommand{\widthh}{0.3\linewidth}
    \setlength{\tabcolsep}{1pt}
    \begin{tabular}{ccccc}
        \includegraphics[width=\width]{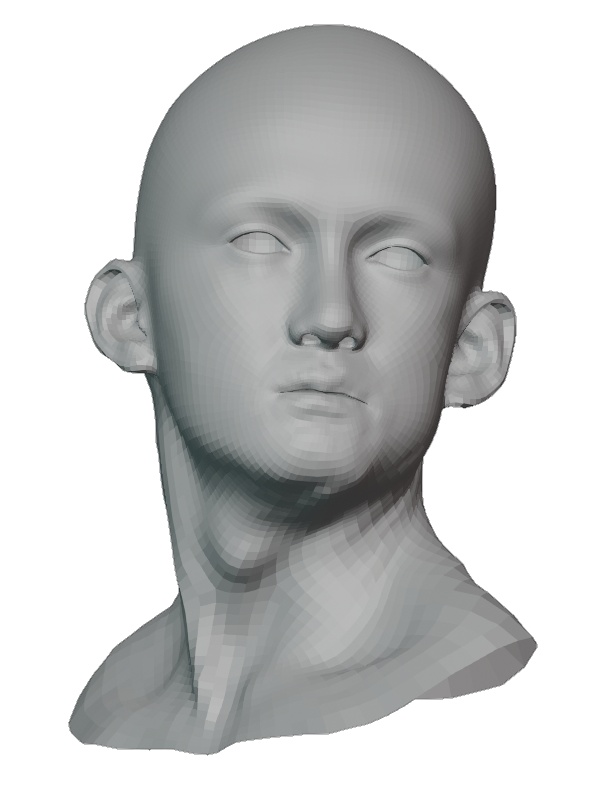} & 
        \includegraphics[width=\width]{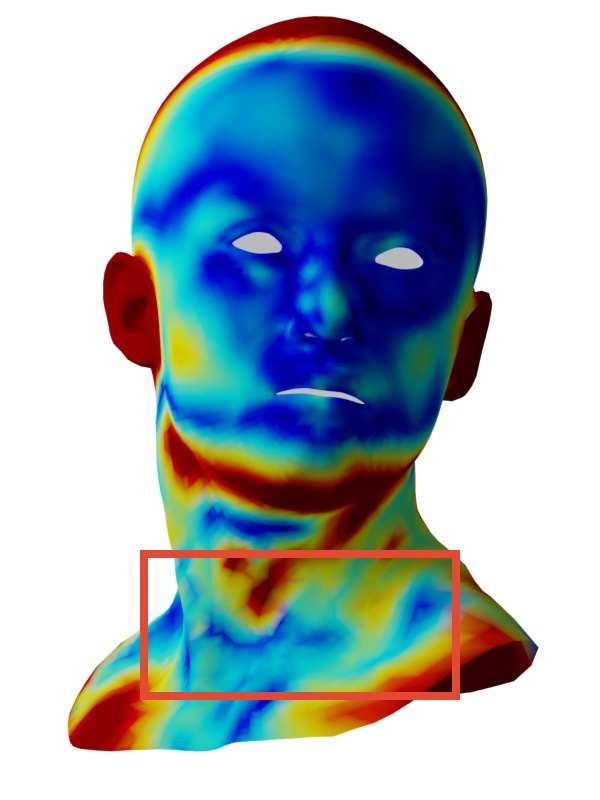} &
        \includegraphics[width=\width]{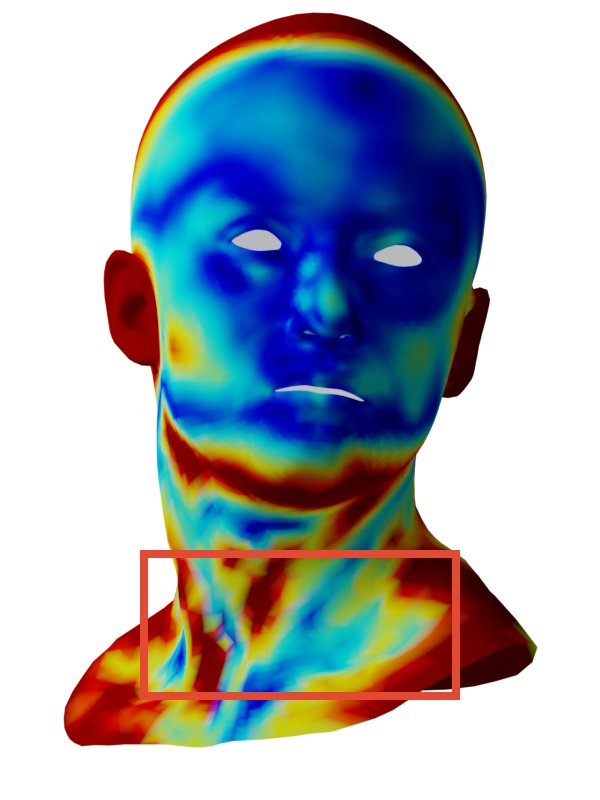} &
        \raisebox{-0cm}{\includegraphics[width=1cm]{fig/realfig/colorbar.png}}
        \\[-0.6cm]
         & 
        \includegraphics[width=\widthh]{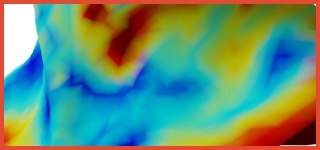} &
        \includegraphics[width=\widthh]{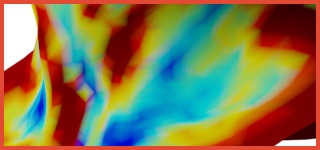}
        \\
        \includegraphics[width=\width]{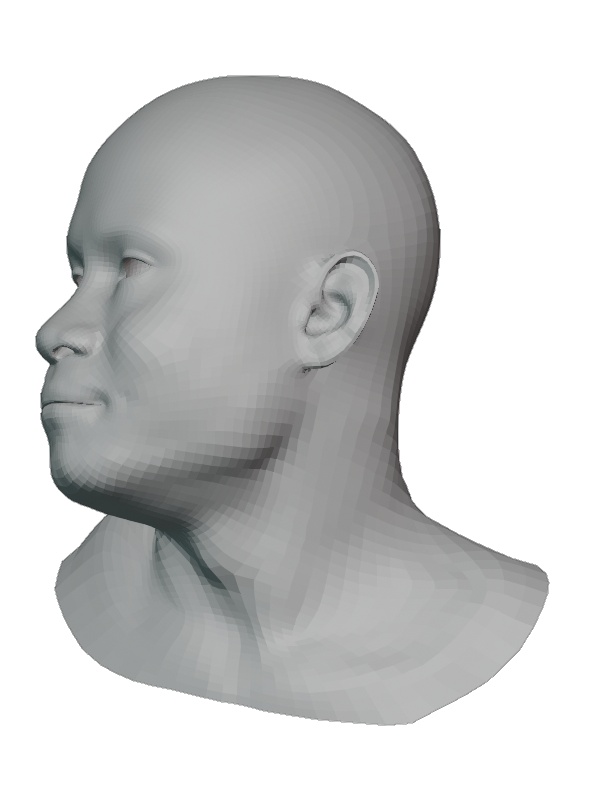} & 
        \includegraphics[width=\width]{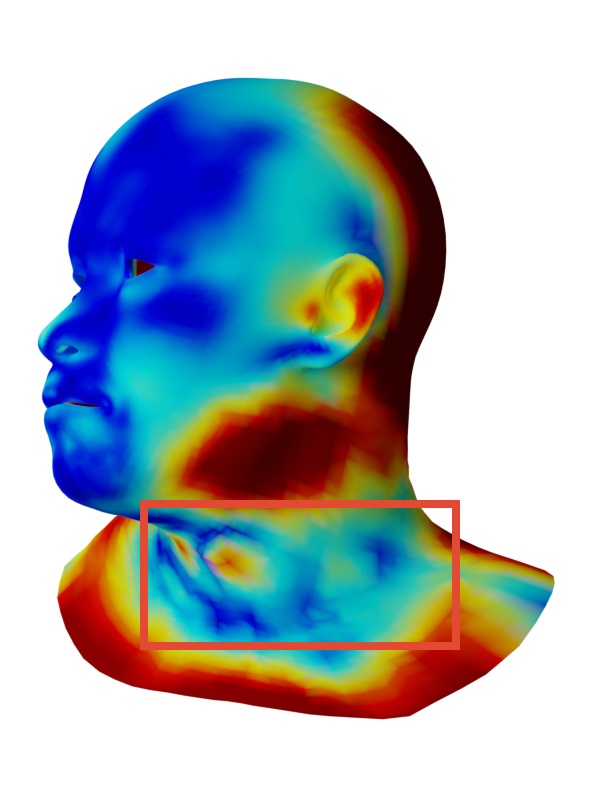} &
        \includegraphics[width=\width]{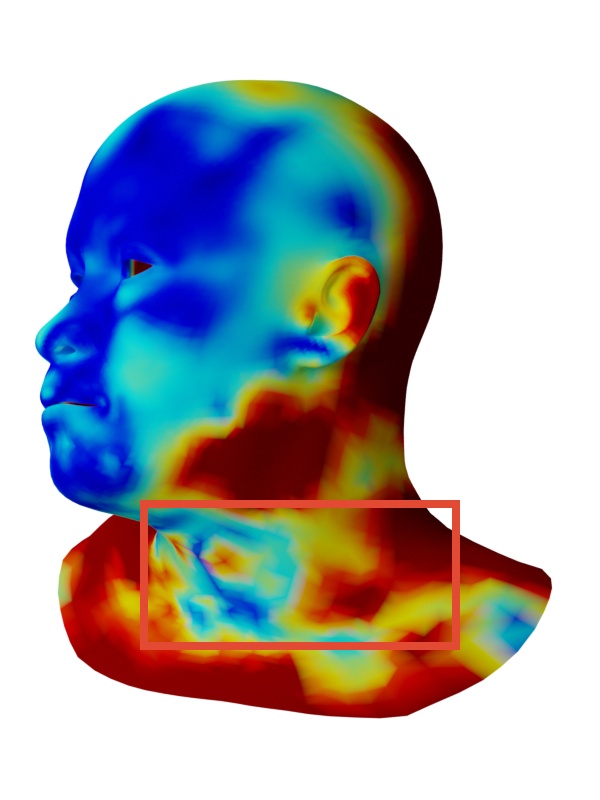} &
        \raisebox{-0cm}{\includegraphics[width=1cm]{fig/realfig/colorbar.png}}
        \\[-0.6cm]
         & 
        \includegraphics[width=\widthh]{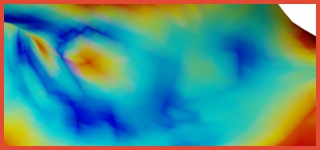} &
        \includegraphics[width=\widthh]{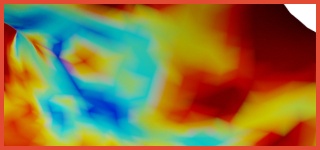}
        \\
        Target & \textbf{personalized $\mathcal{P}$} & \textbf{generic $\mathcal{P}$}
    \end{tabular}
    \caption{Qualitative evaluation of whether to use person-specific pose blendshapes. Using generic pose blendshapes ignores personalized attributes of head and neck movements. For example, performers with extreme body fat rates will result in the Sternocleidomastoid muscle being either stronger prominent bulged (upper) or less defined (lower). With the modeling of personalized pose blendshapes, our model has lower reconstruction error on posed scans.}    \label{fig:ablation}
\end{figure}

To evaluate the generalization of the shape space, larynx shape space, and expression blendshapes space, we randomly divided $90\%$ of the data for training and the remaining $10\%$ for testing. 
Fig.~\ref{fig:quantitative}(a)(b)(c) plot the evaluation result on the generalization ability of the \modelname shape, larynx shape, and expression blendshapes spaces, respectively. 
The generalization error is shown as the mean and standard deviation of RMSE with respect to the number of components. 
The shape space's generalization error decreases monotonically on the testing shape as the number of components increases. The error curve reaches below 0.65 mm and 0.26 mm with 50 and 200 components, respectively. Similarly, the larynx shape space and expression blendshapes space also exhibit decreasing error curves as the number of components increases. 
Fig.\ref{fig:quantitative}(d) shows the reconstruction error of the pose blendshapes space, which decreases as the number of components increases.

\paragraph{Ablation}
\label{sec:ablation}

Modeling the space of expression blendshapes provides fine-grained geometry for different identities under different expressions. In contrast, using generic expression blendshapes ignores the diversity of facial muscles across identities and under different expressions. We denote the original model and the variant one using generic expression blendshapes as \textbf{personalized $\mathcal{E}$} and \textbf{generic $\mathcal{E}$}, respectively. We utilize both models to fit unseen expressions. As illustrated in Fig.~\ref{fig:ablation_E}, modeling person-specific expression blendshapes results in better reconstruction of facial geometry under different expressions, including single and complex expressions.

Similarly, modeling the space of pose blendshapes provides fine-grained geometry for different identities under different poses. In contrast, using generic pose blendshapes ignores person-specific attributes in animation. We denote the original model and the variant one using generic pose blendshapes as \textbf{personalized $\mathcal{P}$} and \textbf{generic $\mathcal{P}$}, respectively. We utilize both models to fit unseen poses. As illustrated in Fig.~\ref{fig:ablation}, modeling person-specific pose blendshapes results in better reconstruction of neck geometry under different poses. One observation is that body fat can obscure the underlying muscle tissue, making the muscles less visible and less defined, and vice versa. Therefore, modeling pose blendshapes space results in better fitting accuracy and higher expressiveness.

\begin{figure}[t]
    \centering
    \includegraphics[width=\linewidth]{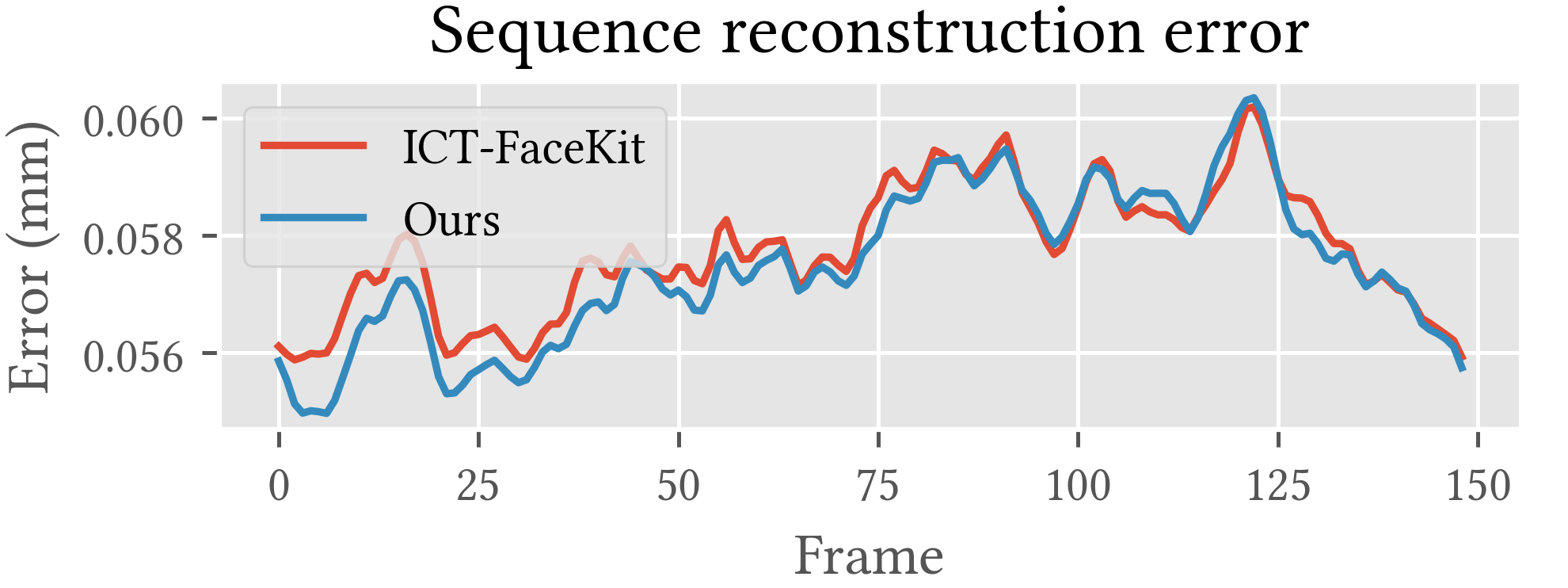}
    \caption{    
Quantitative comparison with ICT-FaceKit on sequence from VOCASET, which uses generic expression blendshapes instead of personalized ones.
    }
    \label{fig:comparison3}
\end{figure}

\begin{figure*}
    \centering
    \newcommand{\width}{0.17\linewidth}
        \includegraphics[width=\linewidth]{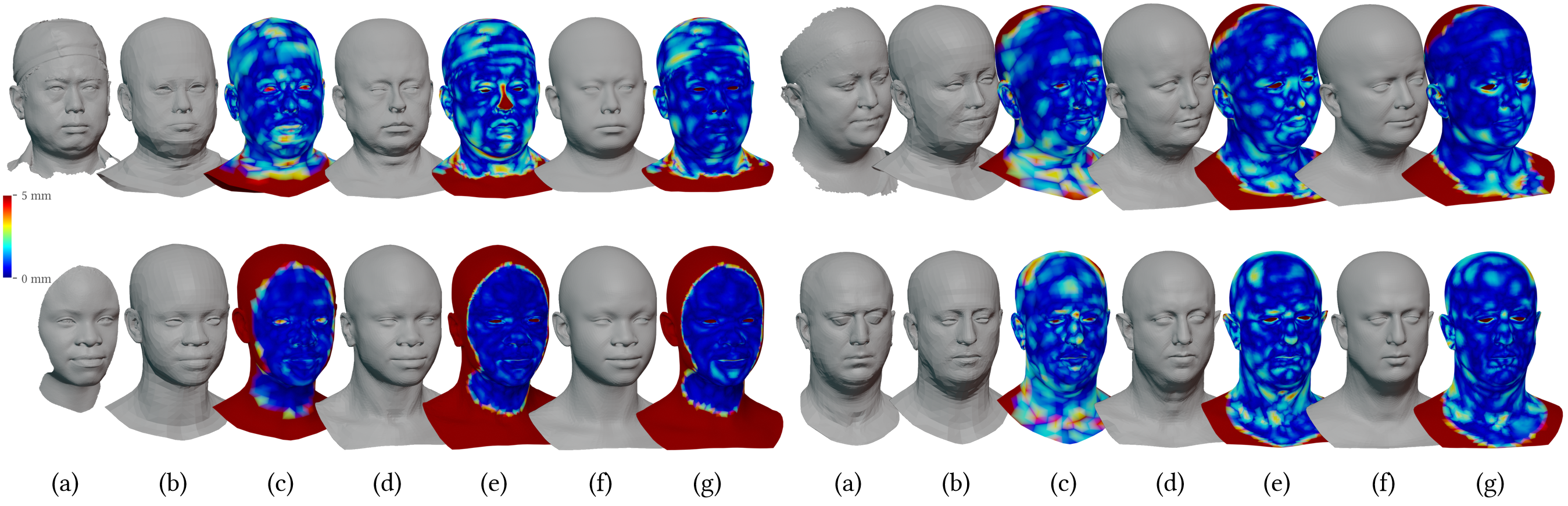}
                  \caption{
Qualitative comparison on registration of samples from FaceScape dataset~\cite{yang2020facescape} (upper left), VOCASET~\cite{VOCA2019} (upper right),  ICT-3DRFE~\cite{stratou2011effect} (lower left) and Multiface~\cite{wuu2022multiface} (lower right).
In the figure, we show (a) the original scan, (b) FLAME registration result, (c) mesh-to-scan distance of FLAME, (d) ICT-FaceKit registration result, (e)  mesh-to-scan distance of ICT-FaceKit, (f) HACK registration result, (g)  mesh-to-scan distance of HACK.
}
    \label{fig:comparison}
\end{figure*}

\begin{figure}[t]
    \centering
    \includegraphics[width=\linewidth]{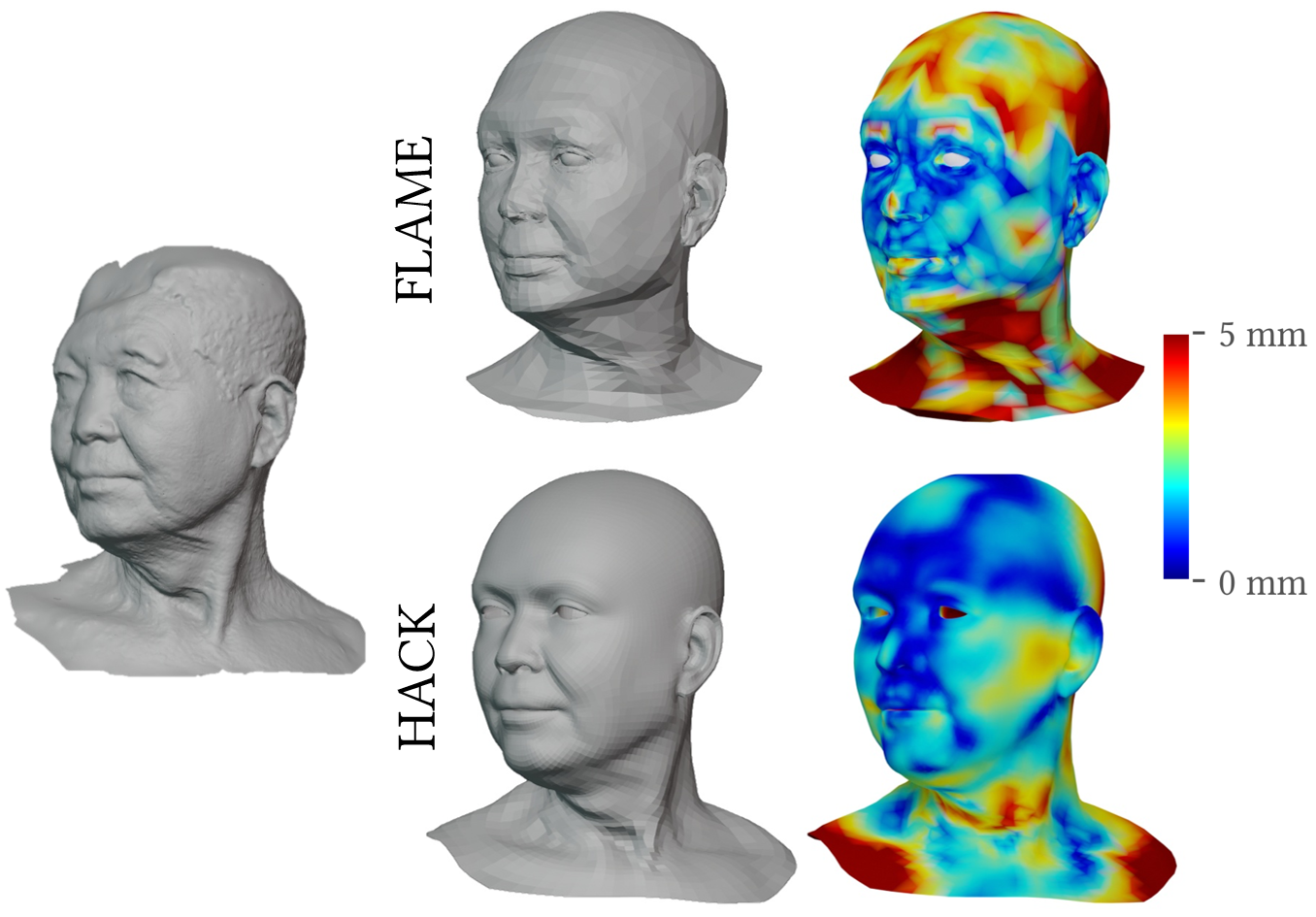}
    \caption{
Qualitative comparison with FLAME on registration of scan from our captured testing data with a head pose.
    }
    \label{fig:comparison2}
\end{figure}

\begin{figure}[t]
    \centering
        \includegraphics[width=\linewidth]{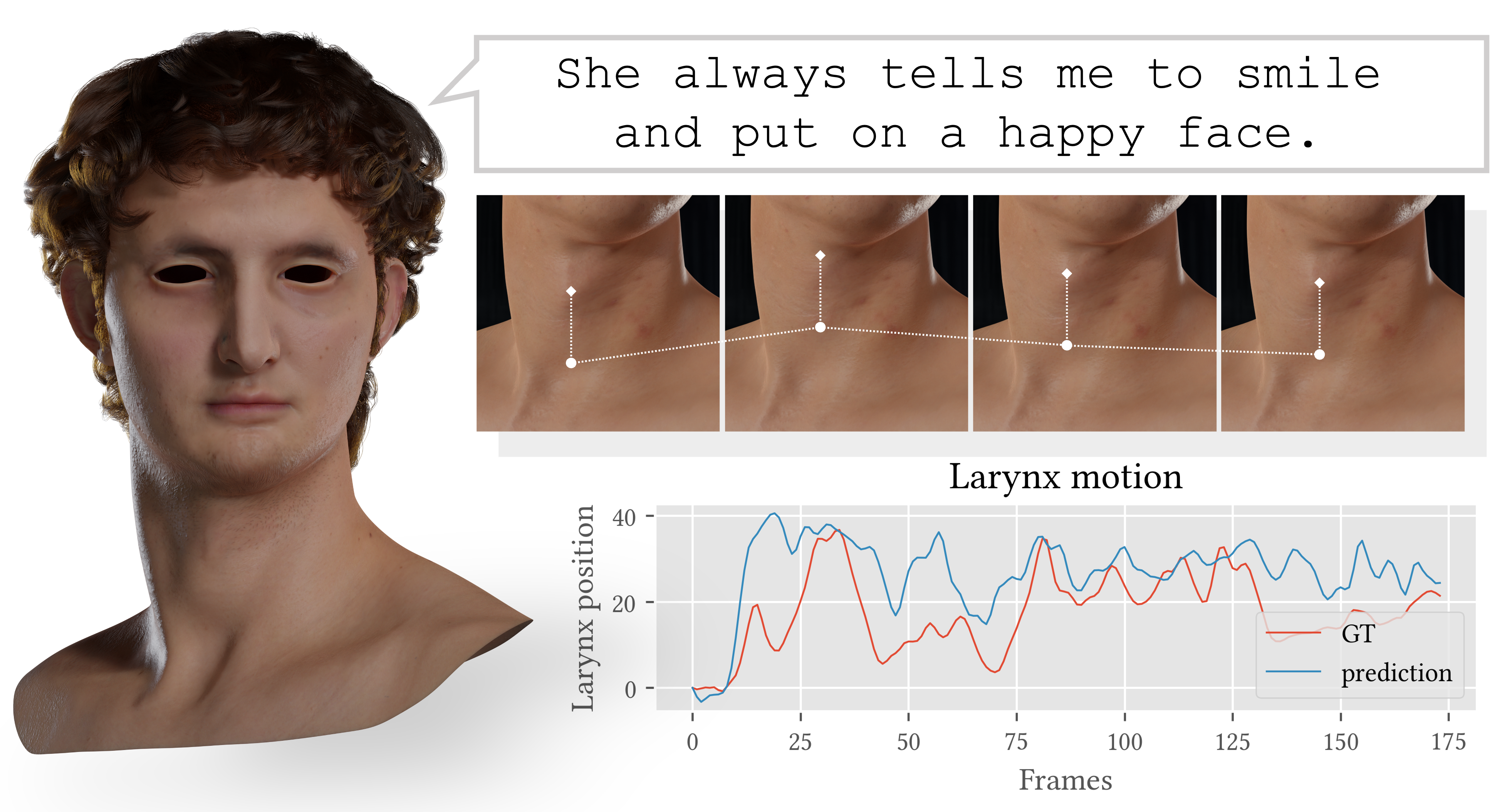}
    \caption{The results of generating larynx motion from expressions when speaking. The performer is speaking the sentence ``She always tells me to smile and put on a happy face". The figures show the larynx position in different frames (upper) and the corresponding generated position sequence (lower).}
    \label{fig:larynx}
\end{figure}

\begin{figure}[t]
    \centering
    \includegraphics[width=\linewidth]{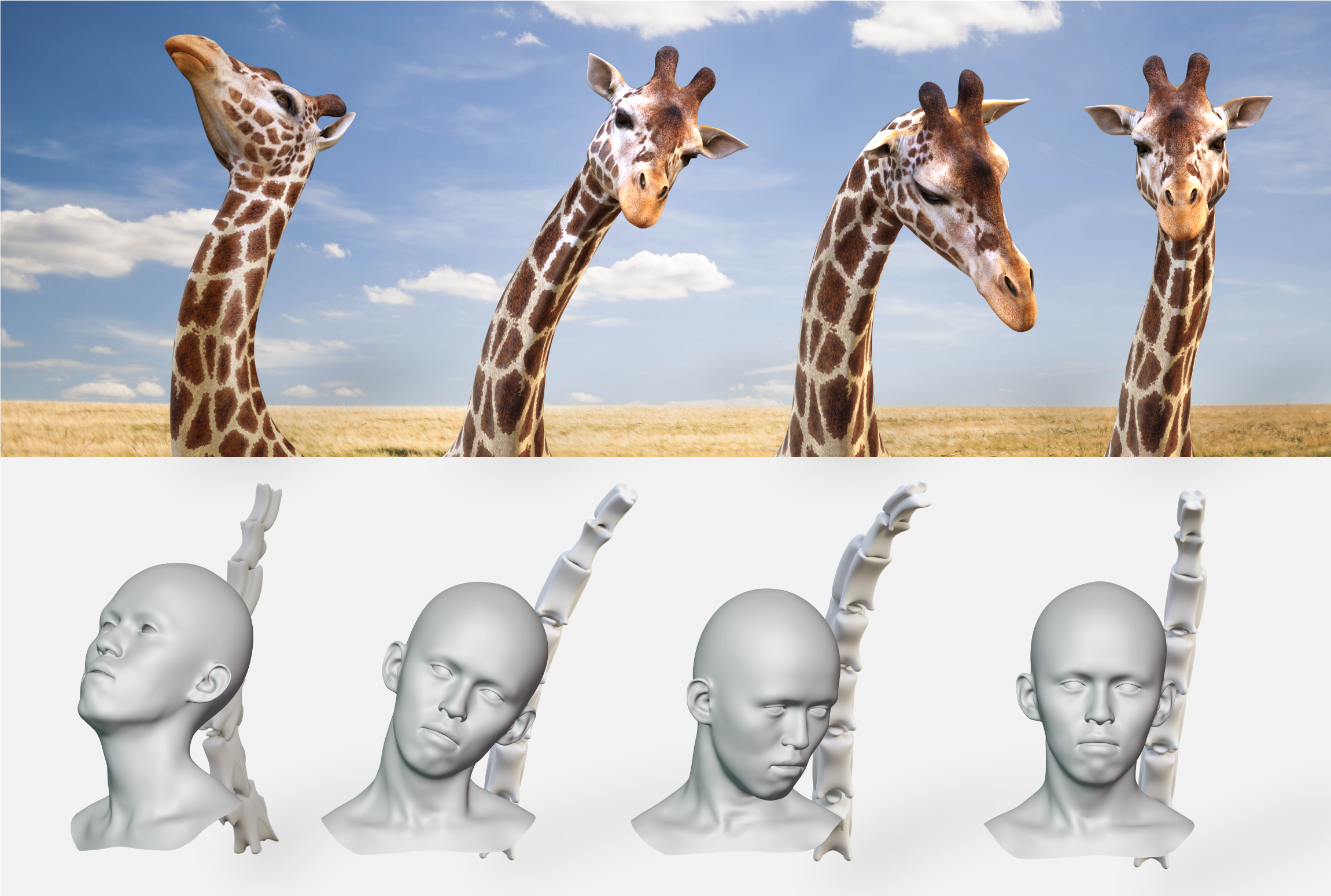}
    \caption{The results of motion transfer from a human to a giraffe. 
    From left to right, we demonstrate different poses including cervical extension, side-bend, flexion, and twisting. The upper figure shows the rendering result with transferred motion, and the lower figure shows the original human pose.
    As discussed in Sec.~\ref{sec:transfer}, \modelname can transfer the motions from one character to another character even if they belong to different species. With \modelname's modeling of the cervical spine, the new character's head and neck movements are transferred naturally and accurately.}    \label{fig:giraffe}
\end{figure}

\subsection{Comparison}
\label{sec:exp_eval}

\paragraph{Quantitative results}

In the task of reconstructing faces in a rest pose, we compared \modelname with two state-of-the-art parametric models, FLAME~\cite{FLAME:SiggraphAsia2017} and ICT-FaceKit~\cite{LearningFaceCVPR2020}, to demonstrate its generalization ability.
We experimented with various identities, including 20 scans from FaceScape~\cite{yang2020facescape}, 12 scans from VOCASET~\cite{VOCA2019}, 22 scans from ICT-3DRFE~\cite{stratou2011effect}, and 10 scans from Multiface~\cite{wuu2022multiface}. To fit identity shape, we used the official \textit{flame-fitting} repository~\cite{FLAME:SiggraphAsia2017}, with the weights of ``scan-to-mesh'', ``landmark'' and ``shape regularization'' setting to $2$, $0.01$ and $0.00005$, respectively. To ensure a fair comparison, we used the first 100 shape components in each model and removed outlier point clouds in the raw scans during optimization.
Table~\ref{tab:comparison} reports the quantitative results of the mean reconstruction error, as the mean and the standard deviation of the average scan-to-mesh distance in millimeters.
In Fig.~\ref{fig:comparison3}, we compared the use of generic expressions from ICT-FaceKit to personalized expressions of \modelname by reconstructing talking sequences from VOCASET.

\paragraph{Qualitative results}

We further compare the qualitative reconstruction results between FLAME, ICT-FaceKit and \modelname.
In Fig.~\ref{fig:comparison}, we reconstruct neutral scans from FaceScape~\cite{yang2020facescape}, VOCASET~\cite{VOCA2019}, ICT-3DRFE~\cite{stratou2011effect}, and Multiface~\cite{wuu2022multiface}.
We provide qualitative registration results and the corresponding mesh-to-scan distance. Note that a regularization term is applied during the fitting process to minimize the fitting objective and avoid artifacts.
In Fig.~\ref{fig:comparison2}, we compare the fitting of posed scans from our captured testing data for both FLAME and \modelname, and provide corresponding error maps.

\vspace{0.5cm}

These comparison results illustrate that our HACK model provides accurate disentanglement for human identity, facial expression, skeletal pose for neck region, and larynx motions. Note that HACK is not intended to outperform previous models in every individual component such as shape or expression. Rather, it provides a comprehensive and anatomically-consistent model for the neck regions. More significantly, we show that by further considering nuanced neck and larynx motions in human parametric modeling, HACK is able to provide personalized and realistic controls for a wide range of applications.

\subsection{Application result}
\label{sec:application_result}

As a parametric model, we can fit the parameters of \modelname to a human head and neck mesh, and further synthesize new expressions, poses, and larynx motions with high fidelity. Existing head modeling techniques lack the ability to accurately estimate neck geometry. To demonstrate this, we fit the parameters of \modelname to the 3D model of Michelangelo's highly regarded sculpture, \textit{David}, which relies on the neck for conveying movement and energy. Using \modelname, we are able to animate \textit{David} with more realism by exploiting HACK's neck modeling ability.
Given a target mesh of \textit{David}, we fit \modelname parameters on it while keeping his unique traits of head and neck using personalized pose blendshapes. As illustrated in the upper right of Fig.~\ref{fig:fitting}, HACK faithfully reconstructs his elegant neck with anatomically-consistent cervical spines. 
We further animate the HACK \textit{David} model. As shown in the upper right of Fig.~\ref{fig:inference}. Notice that with HACK's personalized characteristic modeling ability, \textit{David} preserves personalized traits during animation by contracting his platysma and making a defined bowstring, which demonstrates his strength and strong emotion under different poses, as shown in the upper right of Fig.~\ref{fig:inference}(b).

Larynx motion is an important but also complicated movement, and HACK adopts a fully disentangled design to fully control the appearance and motion of the larynx.
To demonstrate the fine-grained control over larynx motion enabled by HACK, we simulate the swallowing action on the \textit{David} model. Specifically, we extract the corresponding larynx slicing parameter $\tau$ for the swallowing action in the dynamic performance sequence. We align sequences $\tau$ from different clips temporally via peaks and valleys, and then average all sequences to obtain the controlling sequence for the swallowing action. We then apply the obtained $\tau$ sequence on the \textit{David} HACK model to simulate the swallowing action. 
Please refer to the accompanying video for a demonstration of the realistic swallowing simulation.

\paragraph{Larynx motion generation}
As discussed in Sec.~\ref{sec:transformer}, we can utilize a  transformer to generate larynx motion from the expression sequence. We use a state-of-the-art facial expression extractor, ARKit~\cite{arkit}, to obtain the expression sequence $\bm{\Psi}$. The transformer is then applied to $\{\bm{\Psi}\}$, generating the larynx motion sequence $\{\tau\}$.
In Fig.~\ref{fig:larynx}, we show a generated sequence using our testing data, which faithfully restores the larynx motion using given expressions.

\paragraph{Motion transfer}

\modelname can be used to transfer the neck animation from a human to a giraffe, as shown in Fig.~\ref{fig:giraffe}. By transferring the pose parameters between our models, we are able to reproduce realistic neck movements and pose changes in the giraffe, while keeping the skin acting naturally and accurately. Thus, we can generate photo-realistic animations that accurately simulate how a giraffe's neck would move and deform under different conditions.
In comparison to conventional surface models that lack an inner anatomically-prior cervical spine, \modelname is able to achieve more realistic and believable neck animations, as it is able to accurately reproduce the way that the skin would behave and interact with the underlying bones. Please see the animation sequence in the accompanying video for a demonstration of the capabilities of our model.

\subsection{Limitation and Discussion}

HACK provides a compelling parametric disentanglement of the human head and neck, complete with rich inner anatomical structures. However, it still has some limitations. Here, we provide a detailed analysis and discuss potential future extensions.

First, similar to other parametric models~\cite{ploumpis2020towards}, our HACK model could be enhanced by integrating more individual facial components such as eyeballs or teeth~\cite{EyeModel2016,TeethModel2016} for more comprehensive modeling. However, modeling human hair remains challenging and requires more complicated geometry representation such as strands~\cite{FacialHair22}.  
Further combining recent advances in neural rendering (NR)~\cite{NOPC,Lombardi21,wang2022hvh} with HACK could provide photo-realistic hair rendering, but this may sacrifice compatibility with existing CG production pipelines. Furthermore, building a hybrid model that combines both parametric and NR modeling for various components of human characters remains an unsolved research direction with enormous potential. Nevertheless, we believe that our publicly available HACK model will serve as a solid foundation for future explorations
Besides, HACK aims to strike a delicate balance between general and person-specific characteristics within the framework of parametric models. As a result, we achieve more personalized pose and expression results than previous general models such as FLAME~\cite{FLAME:SiggraphAsia2017}. However, it remains extremely challenging to encode more nuanced personalized properties, such as dynamic textures with wrinkles or pore-level details, into the current parametric model while maintaining its convenient controls. Recent NR advances hold promise for synthesizing such personalized details, but the compatibility with CG engines remains an unsolved bottleneck issue.

\section{Conclusion}

We have presented HACK, a novel parametric model for constructing the head and cervical region of digital humans. It significantly models neck and larynx motions, achieving more realistic and anatomically consistent controls for the neck regions that are compatible with CG engines.
Our comprehensive capturing combines 3D ultrasound imaging with multi-view photometric capture system, so as to extract the inner biomechanical structures for the vertebrae of the cervical spine, as well as the external geometry and physically-based textures. HACK separates the depiction of 3D head and neck into various blendshapes on top of the neutral one. Our anatomically-consistent skeletal pose encodes the rich cervical priors, while our expression blenshapes tied to the facial action units enable artist-friendly controls. The adopted larynx blendshapes with both larynx deformation and slicing in the UV-space further provide accurate and nuanced modeling for the larynx beneath the neck skin. We also tackle schemes to augment the pose and expression blendshapes by optimizing an efficient mapping from the identical shape space to the PCA spaces of personalized blendshapes, which significantly improves the personalized characteristics.
We demonstrate the capabilities of HACK model for more realistic and nuanced controls, especially for the neck regions, and showcase the applications using the inter-correlation between head and neck for motion synthesis and transfer.
It makes a solid step for covering more spectrum of parametric digital humans, and hence facilitating numerous potential applications for entertainment, gaming, and immersive experience in VR/AR.

\begin{acks}

This work was supported by National Key R\&D Program of China (2022YFF0902301), NSFC programs (61976138, 61977047), STCSM (2015F0203-000-06), and SHMEC (2019-01-07-00-01-E00003). We also acknowledge support from Shanghai Frontiers Science Center of Human-centered Artificial Intelligence (ShangHAI).

\end{acks}

\bibliographystyle{ACM-Reference-Format}
\bibliography{bibliography}

\end{document}